\documentclass[12pt, a4paper]{article}
 \usepackage[font=small,format=plain,labelfont=bf,up,textfont=normal,up,justification=justified,singlelinecheck=false]{caption}
\usepackage{subcaption}

\usepackage[a4paper, left=2cm,right=2cm]{geometry}
\usepackage[colorlinks=true,linkcolor=black,citecolor=teal,urlcolor=MidnightBlue,filecolor=black]{hyperref}
\usepackage{amsfonts}
\usepackage{amsmath,amssymb}
\usepackage{setspace}
\usepackage[dvipsnames]{xcolor}
\definecolor{SchoolColor}{rgb}{0.6471, 0.1098, 0.1882} 
\usepackage{tensor,cancel,slashed,braket,simplewick}

\usepackage[utf8,applemac]{inputenc}
\usepackage{tensor}
\usepackage{cite}
\usepackage{tikz}
\usepackage{graphicx}
\usepackage{bm} 

\usepackage[utf8,applemac]{inputenc}
\usepackage{tensor}
\usepackage{cite}
\usepackage{tikz}
\usepackage{graphicx}
\newcommand{\bs}{\begin{subequations}}
\newcommand{\es}{\end{subequations}}
\graphicspath{{figure/}}
\usepackage{array}
\usepackage{booktabs} 
\usepackage{multirow}

\usepackage{dcolumn}
\usepackage{bm}

\usepackage{verbatim}

\usepackage{textcomp} 
\usepackage{graphicx} 

\numberwithin{equation}{section}
\newcommand{\bea}{\begin{eqnarray}}
\newcommand{\eea}{\end{eqnarray}}
\newcommand{\be}{\begin{equation}}
\newcommand{\ee}{\end{equation}}
\def\nn{\nonumber}

\def\p{\partial}\def\bz{\bar{z}}

\newcommand{\n}{\nabla}
\newcommand{\beqs}{\begin{eqnarray}}
\newcommand{\eeqs}{\end{eqnarray}}
\numberwithin{equation}{section}

\def\cm{\mathcal{M}}\def\co{\mathcal{O}}\def\ct{\mathcal{T}}\def\cy{\mathcal{Y}}

\def\bz{{\bar{z}}}

\newcommand{\hc}{\mathrm{h.c.}}\def\c.c.{\mathrm{c.c.}}
\def\mn{{\mu\nu}}\def\ab{{\alpha\beta}}\def\rs{{\rho\sigma}}
\def\a{\alpha}\def\b{\beta}\def\g{\gamma}

\def\ep{\epsilon}
\def\th{\theta}

\def\m{\mu}\def\n{\nu}
\def\s{\sigma}

\newcommand{\Rmnum}[1]{\uppercase\expandafter{\romannumeral #1\relax}}

\newcommand{\zxh}{\color{green}}

\begin{document}
\begin{titlepage}

\begin{flushright}\vspace{-3cm}
{\small
\today }\end{flushright}
\vspace{0.5cm}
\begin{center}
	{{ \LARGE{\bf{Quantum flux	operators in higher spin theories}}}}\vspace{5mm}

	\centerline{\large{Wen-Bin Liu\footnote{liuwenbin0036@hust.edu.cn}, Jiang Long\footnote{longjiang@hust.edu.cn} \& Xin-Hao Zhou\footnote{zhouxinhao@hust.edu.cn}}}
	\vspace{2mm}
	\normalsize
	\bigskip\medskip

	\textit{School of Physics, Huazhong University of Science and Technology, \\ Luoyu Road 1037, Wuhan, Hubei 430074, China
	}
	\vspace{25mm}

	\begin{abstract}
		\noindent
		{We construct Carrollian higher spin field theories by reducing the bosonic Fronsdal theories in flat spacetime to future null infinity. We extend the Poincar\'e fluxes to quantum flux operators which generate Carrollian diffeomorphism, namely supertranslation and superrotation. These flux operators form a closed symmetry algebra once including a helicity flux operator which follows from higher spin super-duality transformation. The super-duality transformation is an angle-dependent transformation at future null infinity which generalizes the usual electromagnetic duality transformation. The results agree with the lower spin cases when restricted to $s=0,1,2$.}\end{abstract}

\end{center}

\end{titlepage}
\tableofcontents

\section{Introduction}
Recently, Carrollian manifolds \cite{Une, Gupta1966OnAA} have received much attention due to their relations to null geometries. It has been shown that various physically interesting symmetries could be embedded into the geometric symmetry of Carrollian manifold \cite{Duval_2014a,Duval_2014b,Duval:2014uoa}, including the BMS groups \cite{Bondi:1962px,Sachs:1962wk,Barnich:2009se,Barnich:2010eb,Campiglia:2014yka,Campiglia:2015yka,Freidel:2021fxf}, Newman-Unti group, etc.. Moreover, the
Carrollian diffeomorphism, which preserves the null structure of Carrollian manifolds, is nontrivial \cite{Liu:2022mne, Liu:2023qtr,Liu:2023gwa} since one can construct corresponding quantum flux operators at future null infinity for lower spin ($s=0,1,2$) theories. The quantum flux operators are obtained by analyzing the Poincar\'e flux densities which are radiated to future null infinity. They form a faithful representation of Carrollian diffeomorphism for scalar field theory up to an anomalous term which is the intrinsic central charge of the theory. For massless theories with non-zero helicity, the superrotation calls for super-duality transformation and one should also consider the corresponding helicity flux operators. The results can also be extended to various null hypersurfaces in general dimensions \cite{Li:2023xrr}.

In this paper, we will study the quantum flux operators associated with Carrollian diffeomorphism for higher spin (HS) theories ($s>2$) in four dimensions. Although there is no nontrivial S-matrix for flat space massless HS theories \cite{1964PhRv..135.1049W,Grisaru:1976vm,Aragone:1979hx,Weinberg:1980kq,Porrati:2008rm}, it is still valuable to study the HS theories on null hypersurfaces. At first, while there exist extensions of HS supertranslation and superrotation in the literature \cite{Campoleoni:2017mbt,Campoleoni:2017qot, Campoleoni:2020ejn,Bekaert:2022oeh}, it would be nice to show that the symmetry algebra found in the previous paper \cite{Liu:2023gwa} still remains valid for general spin theories. Indeed, we find a similar helicity flux operator in the HS theory which corresponds to super-duality transformation at the null boundary. Actually, the electromagnetic duality, originating from the exploration of magnetic monopole by Dirac \cite{Dirac:1931kp}, has been extended to various vector theories \cite{tHooft:1974kcl,Polyakov:1974ek,1976PhRvD..13.1592D,Montonen:1977sn}, $p$-form gauge theories \cite{Nepomechie:1984wu,Teitelboim:1985ya,Teitelboim:1985yc
}, gravitational theories \cite{Garcia-Compean:1998ipq,Nieto:1999pn,Hull:2000zn, Casini:2003kf,Henneaux:2004jw,Godazgar:2018qpq,Oliveri:2020xls,Seraj:2022qyt}, supersymmetric theories \cite{Witten:1978mh,Osborn:1979tq,Seiberg:1994rs,Intriligator:1995au} and HS theories \cite{Francia:2004lbf,Bekaert:2002dt,deMedeiros:2003osq,Bekaert:2003az,Boulanger:2003vs,Deser:2004xt, Bunster:2006rt,Henneaux:2016zlu}. The super-duality transformation is an angle-dependent generalization of the usual duality transformation. Secondly, there are consistent interacting HS gauge theories in AdS (dS) spacetime \cite{Vasiliev:1990en,Vasiliev:1992av,Vasiliev:2003ev,Vasiliev:2003cph} and the result in this paper is expected to be valid for more general null hypersurfaces. Third, interacting HS theories in flat spacetime indeed exist \cite{Metsaev:1991mt,Metsaev:1991nb,Ponomarev:2016lrm,Skvortsov:2018jea,Skvortsov:2020wtf,Krasnov:2021nsq,Herfray:2022prf,Adamo:2022lah,Boulanger:2023prx} and this work may provide insight on the analysis of these theories at future null infinity. Finally, the construction of Carrollian HS theories is an interesting topic in its own right. 

The structure of the paper is as follows. 
In section \ref{reviewform}, we will introduce the basic ingredients of the Carrollian manifold and review the coordinate systems we adopt in this article. In section \ref{metric}, we will introduce minimal background on the Fronsdal theory in the flat spacetime. In section \ref{eom}, we will reduce the bulk HS theories to future null infinity and find the boundary equation of motion as well as the symplectic form. We will construct quantum flux operators and compute the Lie algebra they generate in the following section. The helicity flux operator is discussed in section \ref{helicity}. We will conclude in section \ref{conc} and technical details are relegated to several appendices.

\section{Carrollian manifold and coordinate systems}\label{reviewform}
 In this work, we will use the Greek alphabet $\mu,\nu,\rho,\sigma,\lambda,\kappa$ to denote tensor components in Cartesian coordinates. For example, the Minkowski spacetime $\mathbb{R}^{1,3}$ can be described in Cartesian coordinates $x^\mu=(t,x^i)$ 
 \bea 
 ds^2=-dt^2+dx^i dx^i=\eta_{\mu\nu}dx^\mu dx^\nu, 
 \eea where $\mu=0,1,2,3$ denotes the spacetime components and $i=1,2,3$ labels the spatial directions. We will also use the Greek alphabet $\alpha,\beta,\gamma,\delta$ to represent components in retarded coordinates. As an illustration, the metric of the Minkowski spacetime in retarded coordinates $x^\alpha=(u,r,\theta,\phi)$ is 
 \bea 
 ds_{}^2=-du^2-2du dr+r^2\gamma_{AB}d\theta^Ad\theta^B,\quad A,B=1,2.
 \eea The capital Latin alphabet $A,B,\cdots$ will be used to represent the components of tensors on $S^2$ in spherical coordinates.
 The future null infinity $\mathcal{I}^+$ is a three-dimensional Carrollian manifold 
 \bea 
 \mathcal{I}^+=\mathbb{R}\times S^2
 \eea with a degenerate metric 
\bea 
ds_{\mathcal{I}^+}^2\equiv\bm{\gamma}=\gamma_{AB}d\theta^Ad\theta^B \label{degemet}
\eea which could be obtained by choosing a cutoff $r=R$, using a Weyl scaling to remove the conformal factor in the induced metric and taking the limit $R\to\infty$ with the retarded time $u$ fixed.
The spherical coordinates $\theta^A=(\theta,\phi)$ are used to describe the unit sphere whose metric reads explicitly as 
\bea 
\gamma_{AB}=\left(\begin{array}{cc}1&0\\0&\sin^2\theta\end{array}\right).
\eea 
We will also use the notation $\Omega=(\theta,\phi)$ to denote the spherical coordinates in the context. 
The covariant derivative $\nabla_A$ is adapted to the metric $\gamma_{AB}$, while $\nabla_\mu$ adapts to the Minkowski metric in Cartesian frame. The integral measure on $\mathcal{I}^+$ is abbreviated as
\bea 
\int du d\Omega\equiv\int_{-\infty}^\infty du \int_{S^2}d\Omega,
\eea where the integral measure on $S^2$
is 
\bea 
\int d\Omega\equiv \int_{S^2}d\Omega=\int_0^\pi \sin\theta d\theta\int_0^{2\pi}d\phi.
\eea 
The Levi-Civita tensor on $S^2$ is denoted as $\bm\epsilon=\frac{1}{2}\epsilon_{AB}d\theta^A\wedge d\theta^B$ with 
\bea 
\epsilon_{\theta\phi}=-\epsilon_{\phi\theta}=\sin\theta,\quad \epsilon_{\theta\theta}=\epsilon_{\phi\phi}=0.
\eea The Dirac delta function on $S^2$ is 
\be \delta(\Omega-\Omega')=\sin^{-1}\theta \delta(\theta-\theta')\delta(\phi-\phi').
\ee 
Besides the metric \eqref{degemet}, there is also a distinguished null vector 
\be 
\bm\chi=\partial_u
\ee which generates the retarded time direction. The Carrollian diffeomorphism is generated by the vector field 
\bea 
\bm\xi_{f,Y}=f(u,\Omega)\partial_u+Y^A(\Omega)\partial_A
\eea where $f=f(u,\Omega)$ is any smooth function of $\mathcal{I}^+$ while $Y^A=Y^A(\Omega)$ is time-independent and only smooth vector field on $S^2$. The Carrollian diffeomorphism generated by $\bm\xi_f=f(u,\Omega)\partial_u$ is called general supertranslation (GST) while the one generated by $\bm\xi_Y=Y^A(\Omega)\partial_A$ is referred to special superrotation (SSR). 

In the following, we may also use stereographic project coordinates on $S^2$ which are defined by 
\be 
z=\cot\frac{\theta}{2}e^{i\phi},\quad \bar{z}=\cot\frac{\theta}{2}e^{-i\phi}
\ee and the metric of $S^2$ becomes
\be 
\bm\gamma=2\gamma dz d\bar{z},\quad \gamma=\frac{2}{(1+z\bar z)^2}.
\ee 
The volume form reads
\be 
d^2z\equiv -i \gamma dz\wedge d\bar{z}
\ee with the Levi-Civita tensor being
\bea 
\epsilon_{z\bar{z}}=-\epsilon_{\bar z z}=-i\gamma,\quad \epsilon_{zz}=\epsilon_{\bar z\bar z}=0.
\eea The Dirac delta function is defined by
\be 
\delta^{(2)}(z-z')= i\gamma^{-1}\delta(z-z')\delta(\bar z-\bar z').
\ee 
In this coordinate system, any rank $s$ symmetric traceless tensor $T_{A(s)}$ can only have two non-vanishing components 
\bea 
T_{z(s)},\quad T_{\bar{z}(s)}.
\eea Here we use the short notation 
\bea 
T_{A(s)}=T_{(A_1\cdots A_s)}=\frac{1}{s!}\sum_{\pi\in S_s}T_{A_{\pi(1)}A_{\pi(2)}\cdots A_{\pi(s)}}
\eea to represent a rank $s$ symmetric tensor when it causes no confusion.
The element of the permutation group $S_s$ is denoted as $\pi$ in the above equation. The round brackets $(\cdots)$ represent complete symmetrization for the indices inside them. Similarly, the square brackets $[\cdots]$ imply complete antisymmetrization, e.g.,
\bea
T_{[AB]}=\frac{1}{2}(T_{AB}-T_{BA}).
\eea We will also use the abbreviation
\bea 
\nabla_A T_{A(s-1)}\equiv \frac{1}{s}\sum_{i=1}^s\nabla_{A_i}T_{A_1\cdots A_{i-1}A_{i+1}\cdots A_s}=\nabla_{(A_1}T_{A_2\cdots A_s)}
\eea which is a slight abuse of notation. Here the same lower (or upper) indices $A$s are totally symmetrized automatically. One should not be confused with the Einstein summation convention where lower and upper indices are denoted by the same letter.

\section{Metric-like formulation}\label{metric}
In this section, we shall review the metric-like formulation of free massless fields of
arbitrary spin $s$. We shall mainly concentrate
however only on bosonic fields \cite{1978PhRvD..18.3624F} in flat spacetime, while leaving the fermionic HS fields \cite{Fang:1978wz} and HS fields in AdS or dS \cite{Fronsdal:1978vb} spacetime for future study\cite{Fang:1979hq}. As a generalization of the electromagnetism and linearized Einstein gravity, a spin $s$ HS gauge theory ($s>2$) is described by a totally symmetric and doubly traceless Fronsdal field $f_{\mu(s)}$
\bea 
f_{\mu(s)}=f_{\mu_1\cdots\mu_s}=f_{(\mu_1\cdots\mu_s)},\quad f''_{\mu(s-4)}=0
\eea where we use a prime to denote the trace of the HS field
\be 
f'_{\mu(s-2)}=\eta^{\mu(2)}f_{\mu(s)}.
\ee Therefore, a double prime $f''_{\mu(s-4)}$ is the double trace of the HS field 
\be 
f''_{\mu(s-4)}=\eta^{\mu(2)}\eta^{\mu(2)}f_{\mu(s)}.
\ee A totally symmetric rank $s$ field has\footnote{In Appendix \ref{dof}, we review this result in detail.}
\be 
C_{s+d-1}^{d-1}=\frac{(s+d-1)!}{s!(d-1)!}
\ee independent components in general $d$ dimensions. Therefore, the number of independent components of a spin $s$ field is 
\bea 
C_{s+d-1}^{d-1}-C_{s+d-5}^{d-1}.
\eea In four dimensions, this number reduces to $2(1+s^2)$. The spin $s$ field satisfies the Fronsdal equation 
\bea 
\mathcal F_{\mu(s)}\equiv\Box f_{\mu(s)}-s\partial^\nu\partial_{\mu}f_{\mu(s-1)\nu}+\frac12{s(s-1)}\partial_{\mu}\partial_{\mu} f'_{\mu(s-2)}=0
\eea 
which is invariant under the linearized gauge transformation 
\be 
\delta f_{\mu(s)}=s \partial_{\mu}\xi_{\mu(s-1)}
\ee where the rank $s-1$ tensor $\xi_{\mu(s-1)}$ is totally symmetric and traceless
\be 
\xi_{\mu(s-1)}=\xi_{(\mu_1\cdots\mu_{s-1})},\quad \xi'_{\mu(s-3)}=0.
\ee The corresponding action is 
\bea 
S[f]=\int d^4x \mathcal{L}[f]
\eea where the Lagrangian density is 
\bea 
 \mathcal{L}[f]&= -\frac{1}{2}(\partial_\rho f_{\mu(s)})^2 +\frac{1}{2}s\partial_\alpha f_{\beta\mu(s-1)}\partial^\beta f^{\alpha\mu(s-1)}-\frac{1}{2}s(s-1)\partial_\nu f'_{\mu(s-2)} \partial_\rho f^{\nu\rho\mu(s-2)}\nonumber\\
 &+ \frac{1}{4}s(s-1)(\partial_\rho f'_{\mu(s-2)})^2+\frac{1}{8}s(s-1)(s-2)(\partial^\nu f'_{\mu(s-3)\nu})^2.\eea 
The action reduces to the Pauli-Fierz action for $s=2$ and the Maxwell action for $s=1$. The Lagrangian density may be expressed as a compact quadratic form 
\bea 
\mathcal{L}[f]=L^{\rho\mu(s)\sigma\nu(s)}\partial_\rho f_{\mu(s)}\partial_\sigma f_{\nu(s)}
\eea where the rank $2s+2$ tensor $L^{\rho\mu(s)\sigma\nu(s)}$ is symmetric in the index sets $\mu(s)$ and $\nu(s)$ separately. It is also doubly traceless with respect to these two sets of indices
\bea 
L''^{\rho\mu(s-4)\sigma\nu(s)}=0,\quad L''^{\rho\mu(s)\sigma\nu(s-4)}=0.
\eea 
It may be obtained by taking the symmetric and doubly traceless part of the following rank $2s+2$ tensor 
\bea 
 \widetilde{L}^{\mu\mu_1\cdots\mu_s\nu\nu_1\cdots\nu_s}&&={ -\frac{1}{2}\eta^{\mu\nu}\eta^{\mu_1\nu_1}\cdots \eta^{\mu_s\nu_s}+\frac{1}{2}s \eta^{\mu\nu_1}\eta^{\nu\mu_1}\eta^{\mu_2\nu_2}\cdots\eta^{\mu_{s}\nu_s}}\nn\\&&+\frac{1}{4}s(s-1)(-\eta^{\mu\mu_1}\eta^{\nu\mu_2}\eta^{\nu_1\nu_2}-\eta^{\mu\nu_1}\eta^{\nu\nu_2}\eta^{\mu_1\mu_2}+\eta^{\mu\nu}\eta^{\mu_1\mu_2}\eta^{\nu_1\nu_2})\eta^{\mu_3\nu_3}\cdots \eta^{\mu_s\nu_s}\nn\\
 &&+\frac{1}{16}s(s-1)(s-2)(\eta^{\mu\mu_1}\eta^{\nu\nu_1}+\eta^{\mu\nu_1}\eta^{\nu\mu_1})\eta^{\mu_2\mu_3}\eta^{\nu_2\nu_3}\eta^{\mu_4\nu_4}\cdots \eta^{\mu_s\nu_s}
\eea with respect to two sets of indices $\mu(s)$ and $\nu(s)$ separately.

\paragraph{Gauge fixing condition.}
We may choose the following gauge fixing condition 
\bea 
\mathcal G_{\mu(s-1)}\equiv\partial^\nu f_{\mu(s-1)\nu}-\frac{s-1}{2}\partial_{\mu} f'_{\mu(s-1)}=0\label{gf}
\eea to reduce the Fronsdal equation to 
\be 
\partial^2 f_{\mu(s)}=0.\label{boxf}
\ee The gauge fixing condition \eqref{gf} is always possible. More explicitly, we may start from a general field configuration with $\mathcal G_{\mu(s-1)}\not=0$ and choose the gauge parameter $\xi_{\mu(s-1)}$ such that 
\bea 
\mathcal G_{\mu(s-1)}+\partial^\nu\delta f_{\mu(s-1)\nu}-\frac{s-1}{2}\partial_{\mu} \delta f'_{\mu(s-2)}=0.
\eea This is equivalent to the equation 
\be 
\partial^2 \xi_{\mu(s-1)}=-\mathcal{G}_{\mu(s-1)}
\ee whose solution always exists after imposing appropriate initial and boundary conditions. The residue gauge parameter should satisfy the equation 
\be 
\partial^2 \xi_{\mu(s-1)}=0\label{boxxi}
\ee which could be used to set the Fronsdal field to be traceless 
\be 
f'_{\mu(s-2)}=0.\label{gftrace}
\ee This is always possible since the solution of \eqref{boxf} and \eqref{boxxi} is 
\be 
f_{\mu(s)}=\varepsilon_{\mu(s)} (\bm k)e^{ik\cdot x},\quad \xi_{\mu(s-1)}=\kappa_{\mu(s-1)}(\bm k)e^{ik\cdot x},\quad k^2=0
\ee in terms of plane waves. There is no more constraint on the polarization tensors $\varepsilon_{\mu(s)}$ and $\kappa_{\mu(s-1)}$ except that $\kappa_{\mu(s-1)}$ is traceless and $\varepsilon_{\mu(s)}$ is doubly traceless
\be 
\kappa'_{\mu(s-3)}=0,\quad \varepsilon''_{\mu(s-4)}=0.
\ee Considering a solution $f_{\mu(s)}$ which is not traceless
\be 
\varepsilon'_{\mu(s-2)}\not=0,
\ee we may always find a tensor $\kappa_{\mu(s-1)}$ such that 
\bea 
k^{\nu}\kappa_{\nu\mu(s-2)}=-\frac{1}{2}\varepsilon'_{\mu(s-2)}.
\eea 
Therefore, we can always set the HS field to be transverse and traceless. The remaining number of degrees of freedom for the polarization tensor $\varepsilon_{\mu(s)}$ is 
\be 
(C_{s+d-1}^{d-1}-C_{s+d-3}^{d-1})-(C_{s+d-2}^{d-1}-C_{s+d-4}^{d-1})=2s+1,\quad\text{for}\quad d=4.
\ee Similarly, the remaining number of degrees of freedom for the polarization tensor $\kappa_{\mu(s)}$ is $2s-1$. We may impose a further condition 
\be 
n^\nu f_{\nu\mu(s-1)}=0\label{gftrans}
\ee to reduce the number of degrees of freedom to 2. This is the number of propagating degrees of freedom in four dimensions. When we reduce the theory to future null infinity, the fundamental field $F_{A(s)}$ that encodes the radiation information, has exactly two independent components (seeing the next section). The condition \eqref{gftrans} in retarded coordinates becomes 
\be 
f_{r\alpha(s-1)}=0.
\ee 
 
 Such a condition requires 
\begin{align}
 \varepsilon_{r \alpha(s-1)}+sk_{(r}\kappa_{\alpha(s-1))}=0,
\end{align} 
which has $2s-1$ components, and will exhaust the degrees of freedom of $\kappa_{\alpha(s-1)}$.

\section{Asymptotic equation of motion and symplectic form}\label{eom}
Near $\mathcal{I}^+$, we may impose the fall-off condition 
\be 
f_{\mu(s)}=\sum_{k=1}^\infty \frac{F^{(k)}_{\mu(s)}}{r^k}\label{falloff}
\ee for the HS field in Cartesian coordinates. We will abbreviate the leading coefficient as 
\be 
F_{\mu(s)}=F^{(1)}_{\mu(s)}.\label{abr}
\ee The transformations between retarded and Cartesian coordinates are 
\bs\begin{align}
 J^\alpha_{\ \mu}\equiv\frac{\partial x^\alpha}{\partial x^\mu}&=-n_\mu \delta^\alpha_u+m_\mu \delta^\alpha_r-\frac{1}{r}Y_\mu^A\delta_A^\alpha,\\
\bar{J}^\mu_{\ \alpha}\equiv\frac{\partial x^\mu}{\partial x^\alpha}&=\bar{m}^\mu \delta^u_\alpha+n^\mu \delta^r_{\alpha}-r Y^\mu_A \delta^A_\alpha,
\end{align}\es 
where these newly-appearing vectors can be found in Appendix \ref{identities}. The components of the HS field in retarded coordinates can be expressed as 
\be 
f_{\alpha(s)}= \bar{J}^{\mu(s)}_{\ \alpha(s)}f_{\mu(s)}\label{fretarded}
\ee where 
we have used the notation 
\bea 
\bar{J}^{\mu(s)}_{\ \ \alpha(s)}=\bar{J}^{\mu_1}_{\ \alpha_1}\cdots \bar{J}^{\mu_s}_{\ \alpha_s}.\label{notationJ}
\eea 
 By introducing the symbols
\bs\begin{align} 
N^\alpha_{\ \mu}&=-n_\mu\delta^\alpha_u+m_\mu\delta^\alpha_r-Y_\mu^A\delta^\alpha_A,\\
\bar{N}^\mu_{\ \alpha}&=\bar{m}^\mu\delta^u_\alpha+{n}^\mu \delta^r_{\alpha}-Y^\mu_A\delta^A_\alpha,
\end{align}\es
we may define an infinite tower of fields $F^{(k)}_{\alpha(s)}$ on $\mathcal{I}^+$ through the relation
\be 
f_{\mu(s)}=\sum_{k=1}^\infty r^{-k}N^{\alpha(s)}_{\ \ \mu(s)}F_{\alpha(s)}^{(k)}\label{cartesianexp}
\ee where $N^{\alpha(s)}_{\ \ \mu(s)}$ used the same convention as \eqref{notationJ}. Similar to \eqref{abr}, we will always denote 
\be 
F_{\alpha(s)}=F^{(1)}_{\alpha(s)}.
\ee 
Combining \eqref{fretarded} with \eqref{cartesianexp}, we find 
\bea 
f_{\alpha(s)}&&= \bar{J}^{\mu(s)}_{\ \ \alpha(s)}N^{\beta(s)}_{\ \ \mu(s)} \sum_{k=1}^\infty r^{-k}F^{(k)}_{ \beta(s)}.
\eea Using the identities in Appendix \ref{identities}, the fall-off conditions \eqref{falloff} are transformed to 
\be 
 f_{A(m) \hat{\alpha}(s-m)}=r^{m}\sum_{k=1}^\infty \frac{F^{(k)}_{A(m) \hat{\alpha}(s-m)}}{r^k}=r^{m-1}F_{A(m) \hat{\alpha}(s-m)}+\mathcal{O}(r^{m-2}),\quad m=0,1,\cdots,s
\ee where the indices $\hat{\alpha}$ may be chosen as $u$ or $r$. Note that when $m=s$, the fall-off condition for the totally angular components is
\be 
f_{A(s)}=r^{s-1}F_{A(s)}+\cdots
\ee which agrees with the lower spin cases $(s=0,1,2)$.

\subsection{Asymptotic expansions of gauge conditions and EOM}
As has been mentioned, we may impose the following gauge conditions 
\bea 
\partial^{\nu}f_{\nu\mu(s-1)}=0,\quad f'_{\mu(s-2)}=0,\quad n^{\nu}f_{\nu\mu(s-1)}=0
\eea for free HS gauge theory without sources. In retarded coordinates, the third condition leads to 
\be 
F^{(k)}_{r\alpha(s-1)}=0,\quad k=1,2,\cdots.
\ee 
Moreover, the traceless condition \eqref{gftrace} becomes
\bea 
-2f_{ur\alpha(s-2)}+f_{rr\alpha(s-2)}+r^{-2}\gamma^{AB}f_{AB\alpha(s-2)}=0\eea and it follows that \bea \gamma^{AB}f_{AB\alpha(s-2)}=0.
\eea This is the traceless condition on the sphere $S^2$ which is equivalent to 
\bea 
\gamma^{AB}\sum_{k=1}^\infty r^{-k}F_{ABC(s-2)}^{(k)}=0\quad\Rightarrow\quad \gamma^{AB}F^{(k)}_{ABC(s-2)}=0,\quad k=1,2,\cdots.
\eea The transverse condition \eqref{gftrans} is 
\bea 
0=\partial^{\nu}f_{\nu\mu(s-1)}=(-n^{\nu} \partial_u+m^{\nu} \partial_r-\frac{1}{r}Y^{\nu}_A \nabla^A)[N^{\alpha(s)}_{\ \ \mu(s-1)\nu}\sum_{k=1}^\infty r^{-k} F^{(k)}_{\alpha(s)}]
\eea where 
\bea 
N^{\alpha(s)}_{\ \ \mu(s-1)\nu}=N^{\alpha(s-1)}_{\ \ \mu(s-1)}N^{\alpha_s}_{\ \nu}.
\eea 
Using the identities which are shown in Appendix \ref{identities}, we find 
\be 
(k-2) N^{\alpha(s-1)}_{\ \ \mu(s-1)}F^{(k)}_{u\alpha(s-1)}+\nabla^A[N^{\alpha(s-1)}_{\ \ \mu(s-1)}F^{(k)}_{A\alpha(s-1)}]=0,\quad k=1,2,\cdots.
\ee By multiplying the inverse tensor $\bar{N}^{\mu(s-1)}_{\ \ \beta(s-1)}$, the above equation becomes 
\bea 
(k-2)F^{(k)}_{u\beta(s-1)}-(s-1)\delta^A_{\beta}F^{(k)}_{\beta(s-2)Au}+\nabla^CF^{(k)}_{C\beta(s-1)}=0.\label{transconst}
\eea 
\begin{enumerate}
 \item For $\beta(s-1)=u(s-1)$, we find 
 \be 
 (k-2)F^{(k)}_{u(s)}=-\nabla^CF^{(k)}_{Cu(s-1)}.
 \ee The components $F^{(k)}_{u(s)} $ are completely fixed by 
 \be 
 F^{(k)}_{u(s)}=-\frac{1}{k-2}\nabla^CF^{(k)}_{Cu(s-1)}
 \ee except for $k=2$. When $k=2$, we have 
 \be \nabla^CF^{(2)}_{Cu(s-1)}=0
 \ee and $F^{(2)}_{u(s)}$ is free. 
 \item In general, $\beta(s-1)=A(m)u(s-1-m)$, 
 the equation \eqref{transconst} leads to \be 
 F^{(k)}_{A(m) u(s-m)}=-\frac{1}{k-2-m}\nabla^C F^{(k)}_{CA(m) u(s-1-m)},\quad m=1,2,\cdots,s-1\ee 
 except for $k=2+m$.
\end{enumerate} Therefore, at least for $k=1$, all the components like $F_{u\alpha(s-1)}$ are either zero or determined by the symmetric and traceless one $F_{A(s)}$.

\paragraph{Asymptotic equation of motion.}
We still need to solve the EOM \eqref{boxf}. From the identity 
\be 
\partial^2=-2\partial_u\partial_r-\frac{2}{r}\partial_u+\partial_r^2+\frac{2}{r}\partial_r+\frac{1}{r^2}\nabla_A\nabla^A, 
\ee we find 
\begin{align}
 &\partial^2f_{\mu(s)}\nn\\
 &=[-2\partial_u\partial_r-\frac{2}{r}\partial_u+\partial_r^2+\frac{2}{r}\partial_r+\frac{1}{r^2}\nabla_A\nabla^A]\left( N^{\alpha(s)}_{\ \ \mu(s)}\sum_{k=1}^\infty r^{-k}F^{(k)}_{\alpha(s)}\right)\\
 &=\sum_{k\ge 1}^\infty r^{-k-1}[2(k-1) N^{\alpha(s)}_{\ \ \mu(s)} \dot{F}^{(k)}_{\alpha(s)}+(k-1)(k-2)N^{\alpha(s)}_{\ \ \mu(s)} {F}^{(k-1)}_{\alpha(s)}+\nabla^A\nabla_A ( N^{\alpha(s)}_{\ \ \mu(s)} {F}^{(k-1)}_{\alpha(s)})].\nn
\end{align} 
This leads to an infinite tower of equations for the boundary fields
\bea 
2(k-1)\dot{F}^{(k)}_{\beta(s)}+(k-1)(k-2)F^{(k-1)}_{\beta(s)}+\bar{N}^{\ \ \mu(s)}_{\beta(s)}\nabla^A\nabla_A ( N^{\alpha(s)}_{\ \ \mu(s)} {F}^{(k-1)}_{\alpha(s)})=0.
\eea 
It is obvious that there is no dynamical equation for the mode with $k=1$ while all the descendants with $k\ge 2$ are determined through the boundary equations after imposing suitable initial conditions.
\subsection{Symplectic form}\label{symform}
We can find the pre-symplectic form from the variation principle
\bea 
\delta S=\int {\rm EOM}+\int (d^3x)_\mu\Theta^\mu
\eea where 
\be 
\Theta^\rho=2L^{\rho\mu(s)\sigma\nu(s)}\delta f_{\mu(s)}\partial_\sigma f_{\nu(s)}.
\ee The symplectic form can be obtained by a further variation 
\be 
\bm\Omega^{\mathfrak{H}}(\delta f;\delta f;f)=2\int_{\mathfrak{H}} (d^3x)_\rho L^{\rho\mu(s)\sigma\nu(s)}\delta f_{\mu(s)}\wedge \partial_\sigma \delta f_{\nu(s)}
\ee where we have chosen a hypersurface $\mathfrak{H}$ to evaluate the symplectic form. The symplectic form at $\mathcal{I}^+$ is the limit 
\bea
\bm\Omega(\delta F;\delta F;F)&&=\lim_{r\to\infty,\ u\ \text{fixed}} \bm\Omega^{\mathfrak{H}_r}(\delta f;\delta f;f)\nn\\&&=2\int du d\Omega m_\rho L^{\rho\mu(s)\sigma\nu(s)}\delta F_{\mu(s)}\wedge (-n_\sigma) \delta\dot{F}_{\nu(s)}\nn\\&&=\int du d\Omega \delta F_{A(s)}\wedge \delta\dot{F}^{A(s)}\eea where $\mathfrak{H}_r$ is the constant $r$ slice.
It follows that the fundamental commutators are
\bs\label{comF}\begin{align}
 [F_{A(s)}(u,\Omega),F_{B(s)}(u',\Omega')]&=\frac{i}{2}X_{A(s)B(s)}\alpha(u-u')\delta(\Omega-\Omega'),\\
 [F_{A(s)}(u,\Omega),\dot F_{B(s)}(u',\Omega')]&=\frac{i}{2}X_{A(s)B(s)}\delta(u-u')\delta(\Omega-\Omega'),\\
 [\dot F_{A(s)}(u,\Omega),\dot F_{B(s)}(u',\Omega')]&=\frac{i}{2}X_{A(s)B(s)}\delta'(u-u')\delta(\Omega-\Omega'),
\end{align}\es where the function $\alpha(u-u')$ is 
\be \alpha(u-u')=\frac{1}{2}[\theta(u'-u)-\theta(u-u')]
\ee and the rank $2s$ tensor $X_{A(s)B(s)}$ is constructed by 
\be 
X_{A(s)B(s)}=\frac{1}{s!}\sum_{\pi\in S_s}\widetilde{X}_{A_1\cdots A_sB_{\pi(1)}\cdots B_{\pi(s)}}-\text{traces}
\ee where 
\bea 
 \widetilde{X}_{A_1\cdots A_sB_1\cdots B_s}=\gamma_{A_1B_1}\cdots\gamma_{A_sB_s}.
\eea It should be symmetric and traceless among the indices of the same letter\footnote{This property will be referred to as doubly symmetric traceless (concerning two sets of indices). We hope it will not cause confusion with the symmetric and doubly traceless Fronsdal field $f_{\mu(s)}$.}
\bea 
X_{A(s)B(s)}=X_{(A_1\cdots A_s)(B_1\cdots B_s)},\quad \gamma^{A_1A_2}X_{A(s)B(s)}=\gamma^{B_1B_2}X_{A(s)B(s)}=0 .
\eea 
The explicit form of $X^{A(s)B(s)}$ is\footnote{We have derived this formula in Appendix \ref{st2}.}
\bea 
X^{A(s)B(s)}=\sum_{p,q=0}^{[s/2]}a(p,q;s)\gamma^{(A_{1}A_{2}}\cdots \gamma^{A_{2p-1}A_{2p}}\widetilde{X}_{p,q}^{A_{2p+1}\cdots A_{s})(B_{2q+1}\cdots B_s}\gamma^{B_1B_2}\cdots\gamma^{B_{2q-1}B_{2q})},\label{Xsym}
\eea 
with the coefficients $a(p,q;s)$ being 
\be 
a(p,q;s)=(-1)^{p+q}\frac{s![2s-2p-2]!!}{2^{p}p!(s-2p)!(2s-2)!!}\frac{s![2s-2q-2]!!}{2^{q}q!(s-2q)!(2s-2)!!}.
\ee 

The commutators \eqref{comF} can also be derived from canonical quantization which we have checked in Appendix \ref{canoquan}. After defining the vacuum $|0\rangle$ through the annihilation operator in the boundary theory, we obtain the correlation functions 
\bs\begin{align}
 &\bra0 F_{A(s)}(u,\Omega) F_{B(s)}(u',\Omega') \ket0=X_{A(s)B(s)}\beta(u-u')\delta(\Omega-\Omega'),\\
 &\bra0 F_{A(s)}(u,\Omega) \dot F_{B(s)}(u',\Omega') \ket0=X_{A(s)B(s)}\frac{\delta(\Omega-\Omega')}{4\pi(u-u'-i\epsilon)},\\
 &\bra0 \dot F_{A(s)}(u,\Omega) F_{B(s)}(u',\Omega') \ket0=-X_{A(s)B(s)}\frac{\delta(\Omega-\Omega')}{4\pi(u-u'-i\epsilon)},\\
 &\bra0 \dot F_{A(s)}(u,\Omega)\dot F_{B(s)}(u',\Omega') \ket0=-X_{A(s)B(s)}\frac{\delta(\Omega-\Omega')}{4\pi(u-u'-i\epsilon)^2},
\end{align}\es 
where the function $\beta(u-u')$ is defined by 
\bea 
\beta(u-u')=\int_0^\infty \frac{d\omega}{4\pi\omega}e^{-i\omega(u-u'-i\epsilon)}.
\eea

\section{Quantum flux operators}
For any conserved current $j^\mu$, one may construct the corresponding flux across a hypersurface $\mathfrak{H}$ through the formula 
\bea 
\mathcal{F}[j]=\int_{\mathfrak{H}} (d^3x)_\mu j^\mu.
\eea 
To find the Poincar\'e fluxes, the conserved current should be chosen as 
\be 
j^\mu_{\bm\xi}=T^{\mu}_{\ \nu}\xi^\nu
\ee where $T^{\mu\nu}$ is the stress tensor of the theory and $\bm\xi$ is the Killing vectors of Minkowski spacetime. To discuss the fluxes radiated to $\mathcal{I}^+$, we may choose constant $r$ slices $\mathfrak{H}_r$ in retarded coordinates and then take the limit $r\to\infty$ while keeping the retarded
time $u$ finite
\bea 
\lim{}\hspace{-0.8mm}_+=\lim_{r\to\infty,\ u \ \text{finite}}.\label{lim+}
\eea It follows that the Poincar\'e fluxes at $\mathcal{I}^+$ are 
\be 
\mathcal{F}_{\bm\xi}=\lim{}\hspace{-0.8mm}_+\int_{\mathfrak{H}_r}(d^3x)_\mu T^\mu_{\ \nu}\xi^\nu.
\ee 
We may read out the flux density operators from the fluxes arrived at $\mathcal{I}^+$ per unit time and per unit solid angle. The quantum flux operators are the (generalized) Fourier transformation of the normal-ordered flux density operators. However, 
the definition of the stress tensor in HS theories is rather subtle. The conserved gauge invariant Bel-Robinson tensor \cite{2018grav.book.....M, 1980PhRvD..21..358D}, a direct generalization of the canonical stress tensor, is not the quantity we sought for $s\ge 2$ since it has $2s$ derivatives. Though there are various discussions on the gauge invariant conserved currents in the literature \cite{Berends:1985xx,Anselmi:1999bb,Konstein:2000bi,Gelfond:2006be,Bekaert:2010hk}, it is believed \cite{Deser:2003rxd} that there is no gauge invariant stress tensor for $s\ge 2$ due to the no-go theorem of Weinberg and Witten \cite{Weinberg:1980kq}. However, there are gauge non-invariant conserved currents, akin to the Landau-Lifshitz pseudotensor \cite{ 1971ctf..book.....L} in general relativity, which give rise to the gauge invariant conserved charges \cite{Smirnov:2013kba}. Nevertheless, we will treat the HS fields as ordinary matter and use the formula 
\bea 
T_{\rho\sigma}=\frac{-2}{\sqrt{-g}}\frac{\delta S}{\delta g^{\rho\sigma}}\label{stressdef}
\eea to obtain the ``stress tensor''. It turns out that this ``stress tensor'' leads to reasonable flux operators at $\mathcal{I}^+$.
\subsection{Fluxes}
Substituting the Fronsdal action into \eqref{stressdef}, we find 
\bea
T_{\rho\sigma}=\frac{-2}{\sqrt{-g}}\frac{\delta S}{\delta g^{\rho\sigma}}\Big|_{g\to \eta}=\eta_{\rho\sigma}\mathcal{L}[f]-2\frac{\partial L^{\lambda\mu(s)\kappa\nu(s)}}{\partial g^{\rho\sigma}}\partial_{\lambda}f_{\mu_(s)}\partial_{\kappa} f_{\nu(s)}\Big|_{g\to \eta}.
\eea With the conditions \eqref{gf} and \eqref{gftrace}, only the first two terms in the Lagrangian density contribute to the stress tensor
\bea 
T_{\rho\sigma}&&=-\frac{1}{2}\eta_{\rho\sigma}\partial_\nu f_{\mu(s)}\partial^\nu f^{\mu(s)}+\frac{s}{2}\eta_{\rho\sigma}\partial_{\nu_1} f_{\nu_2\mu(s-1)}\partial^{\nu_2} f^{\nu_1\mu(s-1)}+\partial_\rho f_{\mu(s)}\partial_\sigma f^{\mu(s)}\nn\\&&+s\partial_\nu f_{\rho\mu(s-1)}\partial^\nu f_\sigma^{\ \mu(s-1)}-s[\partial_\rho f^\nu_{\ \mu(s-1)}\partial_\nu f_\sigma^{\ \mu(s-1)}+(\rho\leftrightarrow\sigma)]\nn\\&&-s(s-1)\partial_{\nu_1} f^{\nu_2}_{\ \rho\mu(s-2)}\partial_{\nu_2} f^{\nu_1\hspace{4pt}\mu(s-2)}_{\ \ \sigma}.
\eea The stress tensor can be expanded asymptotically near $\mathcal{I}^+$
\bea 
T_{\rho\sigma}=\sum_{k=2}^\infty \frac{t^{(k)}_{\rho\sigma}}{r^k},
\eea where the first few orders are 
\bs\begin{align}
 t^{(2)}_{\rho\sigma}&=n_\rho n_\sigma \dot{F}_{A(s)}\dot{F}^{A(s)},\\
 t^{(3)}_{\rho\sigma}&=n_\rho n_\sigma X_1+n_{(\rho}Y_{\sigma)}^{C}X_C+Y_{(\rho}^{B}Y_{\sigma)}^C X_{BC}+\frac{d}{du}X_{\rho\sigma},\label{tmunu3}
\end{align}\es where 
\bea 
X_{C}&&=2\dot{F}_{A(s)}\nabla_CF^{A(s)}+2(F_{CA(s-1)}\nabla_D\dot{F}^{DA(s-1)}-\dot{F}_{CA(s-1)}\nabla_DF^{DA(s-1)})\\&&-2s(s \dot{F}_{uA(s-1)}F_{C}^{\ A(s-1)}+\dot{F}_{A(s)}\nabla^{A}F_C^{\ A(s-1)} +\dot{F}_{uA(s-1)}F_{C}^{\ A(s-1)}-F_{uA(s-1)}\dot{F}_{C}^{\ A(s-1)}),\nn
\eea
 and the explicit form of $X_1,X_{\rho\sigma}$ as well as $X_{BC}$ are not important in this work. For more details on the calculation, we refer to Appendix \ref{appc}.
We may compute the Poincar\'e fluxes generated by Killing vectors $\bm\xi$
\bea 
\mathcal{F}_{\bm\xi}=\lim\!{}_+\int_{\mathfrak{H}_r} (d^3x)_\rho T^{\rho\sigma}\xi_\sigma.
\eea For spacetime translation generator labeled by a constant vector $c^\mu$
\be 
\bm\xi_c=c^\mu\partial_\mu,
\ee we find energy and momentum fluxes
\bea 
\mathcal{F}_{\bm\xi_c}&&=c^\nu \int du d\Omega m^\mu t^{(2)}_{\mu\nu}\nn\\&&=c^\mu \int du d\Omega n_\mu \dot{F}_{A(s)}\dot{F}^{A(s)}.
\eea For the Lorentz transformation generator
\be 
\bm\xi_\omega=\omega^{\mu\nu}(x_\mu\partial_\nu-x_\nu\partial_\mu)\quad\Leftrightarrow\quad \xi_\omega^\sigma=\omega^{\mu\nu} [(r n_\mu+u\bar{m}_\mu)\delta^\sigma_\nu-(r n_\nu+u\bar{m}_\nu)\delta^\sigma_\mu]
\ee with $\omega^{\mu\nu}$ being a constant antisymmetric tensor, the angular momentum and center-of-mass fluxes are 
\bea 
\mathcal{F}_{\bm\xi_\omega}&&=\lim\!{}_+\int_{\mathfrak{H}_r} du d\Omega m^\rho T_{\rho\sigma}\xi_\omega^\sigma\nn\\&&=\omega^{\mu\nu}\int du d\Omega um^\rho(\bar m_\mu\delta_\nu^\sigma-\bar{m}_\nu\delta_\mu^\sigma) t^{(2)}_{\rho\sigma}+\omega^{\mu\nu}\int du d\Omega m^\rho (n_\mu\delta^\sigma_\nu-n_\nu \delta^\sigma_\mu)t_{\rho\sigma}^{(3)}\nn\\&&=\omega^{\mu\nu}\int du d\Omega \frac{u}{2}\nabla_A Y^A_{\mu\nu}\dot{F}_{A(s)}\dot{F}^{A(s)}-\frac{1}{2}\omega^{\mu\nu}\int du d\Omega Y^A_{\mu\nu}X_A.
\eea 
At the second line, we decompose $t_{\rho\sigma}^{(3)}$ as \eqref{tmunu3}. The total derivative term containing $X_2$ has no contribution after integration by parts. The terms proportional to $n_\rho n_\sigma$ or $Y_{(\rho}^BY_{\sigma)}^C$ are also vanishing due to the identities
\bs\begin{align}
&m^\rho (n_\mu\delta_\nu^\sigma-n_\nu \delta_\mu^\sigma)n_\rho n_\sigma=0,\\
& m^\rho Y_\rho^A=0.
\end{align}\es Using the relation 
\be 
F_{uA(s-1)}=\frac{1}{s}\nabla^CF_{CA(s-1)},
\ee the angular momentum and center-of-mass fluxes become 
\bea 
\mathcal{F}_{\bm\xi_\omega}&&=\omega^{\mu\nu}\int du d\Omega \frac{u}{2}\nabla_CY_{\mu\nu}^C \dot{F}_{A(s)}\dot{F}^{A(s)}\nn\\&&-\omega^{\mu\nu}\int du d\Omega Y_{\mu\nu}^D[\dot{F}_{A(s)}\nabla_DF^{A(s)}-s(\dot F_{CA(s-1)}\nabla^CF_D^{\ A(s-1)}-\dot{F}_D^{\ A(s-1)}\nabla^CF_{CA(s-1)})]. \nn\\
\eea From the Poincar\'e fluxes, we find the following two flux density operators 
\bs\begin{align}
 T(u,\Omega)&=\ :\dot{F}_{A(s)}\dot{F}^{A(s)}:,\\
 M_A(u,\Omega)&=\frac{1}{2}P_{AB(s)CD(s)}(:\dot{F}^{B(s)}\nabla^CF^{D(s)}-{F}^{B(s)}\nabla^C\dot{F}^{D(s)}:).
\end{align}
\es 
The tensor $P_{AB(s)CD(s)}$ is doubly symmetric traceless
\bs\label{doubleystP}\begin{align}
&P_{AB(s)CD(s)}=P_{A(B_1\cdots B_s)CD_1\cdots D_s}=P_{AB_1\cdots B_sC(D_1\cdots D_s)},\\ &P_{AB(s)CD(s)}\gamma^{B(2)}=P_{AB(s)CD(s)}\gamma^{D(2)}=0
\end{align}\es 
and can be obtained from the following tensor 
\bea 
\widetilde{P}_{AB_1\cdots B_sCD_1\cdots D_s}&&=(\gamma_{AC}\gamma_{B_1D_1}+s\gamma_{AB_1}\gamma_{CD_1}-s\gamma_{AD_1}\gamma_{CB_1})\gamma_{B_2D_2}\cdots\gamma_{B_sD_s}.
\eea We have discussed this tensor extensively in Appendix \ref{st2}.
We have added the normal ordering symbol $:\cdots:$ to remove the annihilation operators to the right-hand side of the creation operators. Similar to the lower spin cases, two smeared quantum flux operators can be defined as 
\bs\label{qfo}\begin{align}
 \mathcal{T}_f&=\int du d\Omega f(u,\Omega)T(u,\Omega),\\
 \mathcal{M}_Y&=\int du d\Omega Y^A(u,\Omega)M_A(u,\Omega),
\end{align}\es
where the function $f$ and vector $Y^A$ can be time and angle-dependent. 

\subsection{Supertranslations and superrotations}
The commutators between the quantum flux operators \eqref{qfo} and the fundamental field $F_{A(s)}$ are
\bs\begin{align}
 [\mathcal{T}_f, F_{A(s)}(u,\Omega)]&=-if(u,\Omega)\dot F_{A(s)}(u,\Omega),\\
[\mathcal{M}_Y,F_{A(s)}(u,\Omega)]&=-i\Delta_{A(s)}(Y;F;u,\Omega)+\frac{i}{2}\int du'\alpha(u'-u)\Delta_{A(s)}(\dot Y;F;u',\Omega),
\end{align}\es where 
\bea 
\Delta_{A(s)}(Y;F;u,\Omega)=Y^D\nabla^CF^{B(s)}\rho_{DB(s)CA(s)}+\frac{1}{2}\nabla^CY^D F^{B(s)}P_{DB(s)CA(s)}.
\eea The rank $2s+2$ tensor $\rho_{AB(s)CD(s)}$ is 
\bea 
\rho_{AB(s)CD(s)}=\frac{1}{2}(P_{AB(s)CD(s)}+P_{AD(s)CB(s)})=\gamma_{AC}X_{B(s)D(s)}.
\eea 
After integration by part, the quantum flux operator $\mathcal{M}_Y$ can be rewritten as 
\bea 
\mathcal{M}_Y=\int du d\Omega :\dot{F}^{A(s)}(u,\Omega)\Delta_{A(s)}(Y;F;u,\Omega):.
\eea 
When the test functions $f$ and $Y^A$
are time-independent, the quantum flux operators can be interpreted as supertranslation and superrotation generators. In the literature, the supertranslation and superrotation vectors $\bm\xi_{f,Y}$ are expanded as
\bs\begin{align}
 &\bm\xi_f=f\partial_u+\frac{1}{2}\nabla_A\nabla^A f\partial_r-\frac{\nabla^A f}{r}\partial_A+\cdots,\\
 &\bm\xi_Y=\frac{1}{2}u\nabla_AY^A\partial_u-\frac{1}{2}r\nabla_AY^A\partial_r+\frac{u }{4}\nabla_C \nabla^C\nabla\cdot Y\partial_r+(Y^A-\frac{u}{2r}\nabla^A\nabla\cdot Y)\partial_A+\cdots.
\end{align}
\es in asymptotically flat spacetime. The Lie derivative of the spin $s$ field along the direction of $\bm \xi_{f,Y}$ is 
\bea 
\mathcal{L}_{\bm\xi}f_{\mu(s)}=\xi^\rho\partial_\rho f_{\mu(s)}+ s\partial_{\mu}\xi^\rho\ f_{\mu(s-1)\rho}.
\eea We can read out the variations of the fundamental field under supertranslation and superrotation from the leading order of the components $f_{A(s)}$ as 
\bs\begin{align}
 \delta_f F_{A(s)}&=f\dot{F}_{A(s)},\label{st}\\
 \delta_Y F_{A(s)}&=\frac{1}{2}u \nabla_BY^B \dot{F}_{A(s)}-\frac{1}{2}(s-1)\nabla_BY^B F_{A(s)}+Y^B\nabla_BF_{A(s)}+sF_{A(s-1)C}\nabla_{A}Y^C.\label{sr}
\end{align}\es For the supertranslation of the field $F_{A(s)}$, we find 
\bea 
\delta_f F_{A(s)}=i[\mathcal{T}_f,F_{A(s)}].
\eea We conclude that the quantum flux operator $i\mathcal{T}_f$ is the generator of supertranslation for $f$ being time-independent. For the superrotation of the field $F_{A(s)}$, we should replace the variation \eqref{sr} induced by Lie derivative with the covariant variation \cite{Liu:2023qtr,Liu:2023gwa}
\bea 
\slashed\delta_YF_{B(s)}&=\delta_YF_{B(s)}-s\Gamma^A_{\ B}(Y)F_{AB(s-1)}
\eea where the connection is a symmetric traceless tensor 
\bea 
\Gamma_{AB}(Y)=\frac{1}{2}\Theta_{AB}(Y)=\frac{1}{2}(\nabla_AY_B+\nabla_BY_A-\gamma_{AB}\nabla_CY^C).
\eea After some algebra, we find 
\bea
\slashed\delta_YF_{A(s)}=i[\mathcal{M}_Y,F_{A(s)}]+i[\mathcal{T}_{f=\frac{1}{2}u\nabla_CY^C},F_{A(s)}]
\eea for $Y^A$ being time-independent. In this case, after subtracting a term related to supertranslation, the quantum flux operator $i\mathcal{M}_Y$ should be regarded as the generator of superrotation. 
As a consistency check, one can show that $\ct_f$ and $\cm_Y$ may also be derived from the Hamilton equation $\delta H_{\bm \xi}=i_{\bm \xi}\Omega$ using the above variations.

In \cite{Liu:2022mne,Liu:2023qtr,Liu:2023gwa}, the supertranslation and superrotation generators have been extended through quantum flux operators by including the time dependencies for the functions $f$ and vectors $Y$. However, closing the algebra requires $\dot Y=0$ and then we realize the Carrollian diffeomorphism (intertwined with super-duality transformation), which will be shown in the next subsection for the higher spin theory. It has also been extended in general dimensions and general null hypersurfaces in \cite{Li:2023xrr}.

\subsection{The algebra among flux operators}
Now it is straightforward to compute the commutators for the quantum flux operators 
\bs\begin{align}
 &[\mathcal{T}_{f_1},\mathcal{T}_{f_2}]={\rm C}_T(f_1,f_2)+i\mathcal{T}_{f_1\dot f_2-f_2\dot f_1},\\
 &[\mathcal{T}_f,\mathcal{M}_Y]=-i\mathcal{T}_{Y^A\nabla_A f}+i\mathcal{M}_{f\dot Y}+\frac{i}{2}s\mathcal{O}_{\dot Y^A\nabla^B f \epsilon_{BA}}+\frac{i}{4}\mathcal{Q}_{\frac{d}{du}(\dot Y^A\nabla_A f)},\\
 &[\mathcal{M}_Y,\mathcal{M}_Z]={\rm C}_M(Y,Z)+i\mathcal{M}_{[Y,Z]}+is\mathcal{O}_{o(Y,Z)}+{\rm N}_M(Y,Z),\\ &[\mathcal{T}_f,\mathcal{O}_g]=i \mathcal{O}_{f\dot g},\\
 &[\mathcal{M}_Y,\mathcal{O}_g]={\rm C}_{MO}(Y,g)+i\mathcal{O}_{Y^A\nabla_A g}+{\rm N}_{MO}(Y,g),\\
 &[\mathcal{O}_{g_1},\mathcal{O}_{g_2}]={\rm C}_O(g_1,g_2)+{\rm N}_O(g_1,g_2).
\end{align}\es The results are quite similar to the lower spin cases. We will discuss these commutators term by term.
\begin{enumerate}
\item New local operators. The operator $\mathcal{O}_g$ is 
\bea 
\mathcal{O}_g&&=\int dud\Omega g(u,\Omega):\dot{F}^{DB(s-1)}F^E_{\hspace{4pt}B(s-1)}:\epsilon_{ED}\nn\\
 & &=\int dud\Omega g(u,\Omega):\dot{F}^{DA(s-1)}F^{EB(s-1)}:\epsilon_{ED}\gamma_{A_1B_1}\cdots \gamma_{A_sB_s}\nn\\
 &\equiv &\int dud\Omega g(u,\Omega):\dot{F}^{A(s)}F^{B(s)}:Q_{A(s)B(s)},\label{helicityflux}
\eea where the rank $2s$ tensor $Q_{A(s)B(s)}$ is doubly symmetric traceless
\bea 
Q_{A(s)B(s)}=Q_{(A_1\cdots A_s)(B_1\cdots B_s)},\quad \gamma^{A(2)}Q_{A(s)B(s)}=\gamma^{B(2)}Q_{A(s)B(s)}=0. 
\eea It can be obtained from the tensor \(\epsilon_{B_1A_1}\gamma_{A_2B_2}\cdots \gamma_{A_sB_s}\) using the formula in Appendix \ref{st2}. This operator is the helicity flux operator associated with HS duality transformation which will be discussed in the next section. The other new operator $\mathcal{Q}_h$ is defined as
\bea \mathcal{Q}_h=\int dud\Omega\ h(u,\Omega):{F}^{A(s)}F_{A(s)}:.
\eea Its commutator with the fundamental field $F_{A(s)}$ is non-local and we do not find a physical interpretation for this operator. Therefore, we will not pay more attention to it in the following.
 \item The central terms come from two-point functions for the quantum flux operators 
\bs\begin{align}
 &{\rm C}_T(f_1,f_2)=-\frac{i\delta^{(2)}(0)}{24\pi}\mathcal{I}_{f_1\dddot{f}_2-f_2\dddot{f}_1},\\
 &{\rm C}_M(Y,Z)=\int du du'd\Omega d\Omega' Y^A(u,\Omega)Z^{B'}(u',\Omega')\Lambda_{AB'}^{(s)}(\Omega-\Omega')\eta(u-u'),\\
 &{\rm C}_{MO}(Y,g)=-2s\delta^{(2)}(0)\int du du'd\Omega Y^A(u,\Omega)\nabla^Bg(u',\Omega) \epsilon_{AB}\eta(u-u'),\\
 &{\rm C}_O(g_1,g_2)=4\delta^{(2)}(0)\int du du'd\Omega \eta(u-u')g_1(u,\Omega)g_2(u',\Omega),
\end{align}\es where \begin{align}
 \eta(u-u')= -\frac{\beta(u-u')-\frac{1}{4\pi}}{8\pi(u-u'-i\epsilon)^2}+\frac{\beta(u'-u)-\frac{1}{4\pi}}{8\pi(u'-u-i\epsilon)^2},
\end{align}
and 
\begin{align}
 \Lambda_{AE'}^{(s)}&=P_{AB(s)CD(s)}P_{E'F'(s)G'H'(s)}\nonumber\\
 &\times \big[X^{B(s)F'(s)}\delta(\Omega-\Omega')\nabla^C\nabla^{G'}\left(X^{D(s)H'(s)}\delta(\Omega-\Omega')\right)\nn\\
 &-\nabla^C\left(X^{D(s)F'(s)}\delta(\Omega-\Omega')\right)\nabla^{G'}\left(X^{B(s)H'(s)}\delta(\Omega-\Omega')\right)\big].
\end{align} The identity operator $\mathcal{I}_f$ is defined by
\be 
\mathcal{I}_f=\int du d\Omega \ f(u,\Omega).
\ee The divergence of the Dirac delta function $\delta^{(2)}(0)$ has been regularized to $\frac{1}{12\pi}$ using the Riemann zeta function or heat kernel method \cite{Li:2023xrr}.
\item Non-local terms. The non-local terms are 
\bs \begin{align}
 &{\rm N}_{M}(Y,Z)=\frac{i}{2}\int dudu'd\Omega\alpha(u'-u)\Delta_{A(s)}(\dot Y;F;u')\Delta^{A(s)}(\dot Z;F;u),\\
 &{\rm N}_{MO}(Y,g)=\frac{i}{2}\int dudu'd\Omega \alpha(u'-u)\Delta_{A(s)}(\dot g;F;u)\Delta^{A(s)}(\dot Y;F;u'),\\
 &{\rm N}_O(g_1,g_2)=\frac i2\int dudu'd\Omega\alpha(u'-u)\Delta_{A(s)}(\dot g_1;F;u')\Delta^{A(s)}(\dot g_2;F;u).
 \end{align}
\es Here the tensor $\Delta_{A(s)}(g;F;u)$ is a shorthand of $\Delta_{A(s)}(g;F;u,\Omega)$ and one should distinguish it from $\Delta_{A(s)}(Y;F;u,\Omega)$ which is the superrotation variation of the fundamental field $F_{A(s)}$. Actually, it is defined as 
\bea 
\Delta_{A(s)}(g;F;u,\Omega)=g(u,\Omega) Q_{A(s)B(s)}F^{B(s)}
\eea which relates to the commutator 
\bea 
[\mathcal{O}_g,F_{A(s)}(u,\Omega)]=-i\Delta_{A(s)}(g;F;u,\Omega)+\frac{i}{2}\int du'\alpha(u'-u)\Delta_{A(s)}(\dot g;F;u',\Omega).
\eea 
\item There is a closed algebra for $\dot Y=\dot g=0$ which is similar to the intertwined algebra in the lower spin cases
\bs\label{trun}\begin{align}
 &[\mathcal{T}_{f_1},\mathcal{T}_{f_2}]={\rm C}_T(f_1,f_2)+i\mathcal{T}_{f_1\dot f_2-f_2\dot f_1},\\
 &[\mathcal{T}_f,\mathcal{M}_Y]=-i\mathcal{T}_{Y^A\nabla_A f},\\
 &[\mathcal{M}_Y,\mathcal{M}_Z]=i\mathcal{M}_{[Y,Z]}+is\mathcal{O}_{o(Y,Z)},\label{mymz}\\ &[\mathcal{T}_f,\mathcal{O}_g]=0,\\
 &[\mathcal{M}_Y,\mathcal{O}_g]=i\mathcal{O}_{Y^A\nabla_A g},\\
 &[\mathcal{O}_{g_1},\mathcal{O}_{g_2}]=0.
\end{align}\es This algebra is one of the main results of this paper. The spin $s$ on the right-hand side of \eqref{mymz} can be absorbed into the definition of $\mathcal{O}_g$ and the resulting algebra is isomorphic to each other for $s\not=0$.
\end{enumerate}

\section{Duality transformation}\label{helicity}
In this section, we will confirm that the operator $\mathcal{O}_g$ is the helicity flux operator associated with special super-duality transformation. 

\paragraph{Curvature tensor.}
For a HS field $f_{\mu(s)}$, we may define a curvature tensor \cite{Bunster:2006rt}
\bea 
R_{\mu_1\nu_1\mu_2\nu_2\cdots\mu_s\nu_s}=-2\delta^{\rho_1\sigma_1}_{\mu_1\nu_1}\cdots\delta^{\rho_s\sigma_s}_{\mu_s\nu_s}\partial_{\rho(s)}f_{\sigma(s)}\label{Rf}
\eea where the tensor $\delta^{\alpha\beta}_{\mu\nu} $ is 
\be 
\delta^{\alpha\beta}_{\mu\nu}=\delta^\alpha_\mu\delta^\beta_{\nu}-\delta^\alpha_\nu\delta^\beta_{\mu}
\ee 
and 
\be
\partial_{\rho(s)}=\partial_{\rho_1}\cdots\partial_{\rho_s}.
\ee 
Due to the antisymmetric property of $\delta^{\alpha\beta}_{\mu\nu}$
\bea 
\delta^{\alpha\beta}_{\mu\nu}=-\delta^{\beta\alpha}_{\mu\nu}=-\delta^{\alpha\beta}_{\nu\mu}=\delta^{\beta\alpha}_{\nu\mu},
\eea the curvature tensor is antisymmetric under the exchange of indices $\mu_i$ and $\nu_i$
\be 
R_{\mu_1\nu_1\cdots \mu_i\nu_i\cdots\mu_s\nu_s}=-R_{\mu_1\nu_1\cdots\nu_i\mu_i\cdots\mu_s\nu_s},\quad i=1,2,\cdots,s.
\ee It is also invariant under the exchange of any pair of indices $(\mu_i\nu_i)$ and $(\mu_j\nu_j)$
\be 
R_{\cdots\mu_i\nu_i\cdots\mu_j\nu_j\cdots}=R_{\cdots\mu_j\nu_j\cdots\mu_i\nu_i\cdots},\quad i,j=1,2,\cdots,s.
\ee The cyclic identity 
\bea 
R_{[\mu_1\nu_1\mu_2]\nu_2\cdots}=0
\eea and the Bianchi identity 
\be 
\partial_{[\rho}R_{\mu_1\nu_1]\mu_2\nu_2\cdots}=0
\ee are also satisfied similar to the Riemann tensor.
The Fronsdal equation is equivalent to the vanishing of the ``Ricci'' tensor 
\be 
R_{\mu_1\nu_1\mu_2\nu_2\cdots\mu_s\nu_s}\eta^{\nu_1\nu_2}=0.
\ee 
The dual of the curvature tensor is defined through the Levi-Civita tensor 
\be 
\widetilde{R}_{\mu_1\nu_1\mu_2\nu_2\cdots\mu_s\nu_s}=-\frac{1}{2}\epsilon_{\mu_1\nu_1\rho\sigma}R^{\rho\sigma}_{\ \ \mu_2\nu_2\cdots\mu_s\nu_s}
\ee and has the same symmetry as the curvature tensor. It also obeys the Bianchi identity 
\be 
\partial_{[\rho}\widetilde{R}_{\mu_1\nu_1]\mu_2\nu_2\cdots\mu_s\nu_s}=0
\ee and satisfies the equation of motion 
\be 
\widetilde{R}_{\mu_1\nu_1\mu_2\nu_2\cdots\mu_s\nu_s}\eta^{\nu_1\nu_2}=0.
\ee 

\paragraph{Duality transformation and the corresponding flux.}
The duality transformation is a rotation between the curvature tensor and its dual 
\bs\begin{align}
 R_{\mu_1\nu_1\cdots\mu_s\nu_s}&\to R_{\mu_1\nu_1\cdots\mu_s\nu_s}\cos\varphi+\widetilde{R}_{\mu_1\nu_1\cdots\mu_s\nu_s}\sin\varphi,\\ \widetilde{R}_{\mu_1\nu_1\cdots\mu_s\nu_s}&\to -R_{\mu_1\nu_1\cdots\mu_s\nu_s}\sin\varphi+\widetilde{R}_{\mu_1\nu_1\cdots\mu_s\nu_s}\cos\varphi
\end{align}\es
with $\varphi$ a constant angle. We may introduce a dual Fronsdal field $\widetilde{f}_{\mu(s)}$ which has the same symmetry as the Fronsdal field and relate it to the dual curvature tensor 
\be 
\widetilde{R}_{\mu_1\nu_1\mu_2\nu_2\cdots\mu_s\nu_s}=-2\delta^{\rho_1\sigma_1}_{\mu_1\nu_1}\cdots\delta^{\rho_s\sigma_s}_{\mu_s\nu_s}\partial_{\rho(s)}\widetilde{f}_{\sigma(s)}.
\ee Thus the duality transformation may be induced by rotating the fields $f$ and $\widetilde{f}$
\bs\label{hsduality}\begin{align}
 f'_{\mu(s)}&=f_{\mu(s)}\cos\varphi+\widetilde{f}_{\mu(s)}\sin\varphi,\\
 \widetilde{f}'_{\mu(s)}&=-f_{\mu(s)}\sin\varphi+\widetilde{f}_{\mu(s)}\cos\varphi
\end{align}\es whose infinitesimal transformations are 
\be 
\delta_{\epsilon}f_{\mu(s)}=\epsilon\widetilde{f}_{\mu(s)},\quad \delta_{\epsilon}\widetilde{f}_{\mu(s)}=-\epsilon f_{\mu(s)}
\ee with $\epsilon$ a small positive parameter.

Similar to the vector and gravitational cases, we introduce a symmetric Fronsdal action 
\be 
S[f,\widetilde{f}]=\frac{1}{2}(S[f]+S[\widetilde{f}]).
\ee There is a parallel dual gauge transformation generated by a symmetric traceless tensor $\widetilde{\xi}_{\mu(s-1)}$
\be 
\delta\widetilde{f}_{\mu(s)}=s\partial_{(\mu_1}\widetilde{\xi}_{\mu_2\cdots\mu_s)}.
\ee From Noether's theorem, we can find a conserved current associated with the global duality transformation 
\bea 
j^\rho_{\text{duality}}&&=\frac{1}{2}\frac{\partial\mathcal{L}[f]}{\partial \partial_\rho f_{\mu(s)}}\delta_\epsilon f_{\mu(s)}+\frac{1}{2}\frac{\partial\mathcal{L}[\widetilde{f}]}{\partial \partial_\rho \widetilde{f}_{\mu(s)}}\delta_\epsilon \widetilde{f}_{\mu(s)}\nn\\
&&=L^{\rho\nu(s)\sigma\mu(s)}(\widetilde{f}_{\nu(s)}\partial_\sigma f_{\mu(s)}-f_{\nu(s)}\partial_\sigma \widetilde{f}_{\mu(s)}).
\eea
In the last step, we have omitted the constant parameter $\epsilon$.
We may expand the dual Fronsdal field as 
\be 
\widetilde{f}_{\mu(s)}=\sum_{k=1}^\infty r^{-k} N^{\alpha(s)}_{\ \ \mu(s)}\widetilde{F}_{\alpha(s)}^{(k)}
\ee near $\mathcal{I}^+$ and impose the gauge fixing conditions
\bea 
\partial^{\nu}\widetilde{f}_{\nu\mu(s-1)}=0,\quad \widetilde{f}'_{\mu(s-2)}=0,\quad n^{\nu}\widetilde{f}_{\nu\mu(s-1)}=0.
\eea
Then the helicity flux which radiates to $\mathcal{I}^+$ is 
\be 
\lim\!{}_{+}\int_{\mathfrak{H}_r}(d^3x)_\mu j_{\text{duality}}^\mu=\int du d\Omega \widetilde{F}_{A(s)}\dot{F}^{A(s)}=\int du d\Omega \dot{F}^{A(s)}Q_{A(s)B(s)}F^{B(s)}.
\ee We can read out the helicity density operator 
\be 
O(u,\Omega)=\ :\dot{F}^{A(s)}Q_{A(s)B(s)}F^{B(s)}:
\ee and construct the helicity flux operator 
\bea 
\mathcal{O}_g=\int du d\Omega g(u,\Omega)O(u,\Omega).
\eea This operator is exactly the same as \eqref{helicityflux}. According to the terminology of \cite{Liu:2023gwa}, it becomes the generator of duality transformation for $g=\text{const.}$ and generates special super-duality transformation when $g=g(\Omega)$. 

\paragraph{Why helicity flux?}

Now let us show why we call $\mathcal{O}_g$ helicity flux operator by substituting the mode expansion of the fundamental field in Appendix \ref{canoquan} into $\mathcal{O}_g$. We focus on the special case $g=1$
\begin{align}
 \co_{g=1}&=\int dud\Omega Q^{A(s)A'(s)}\int_0^\infty \frac{d\omega}{\sqrt{4\pi\omega}}\int_0^\infty \frac{d\omega'}{\sqrt{4\pi\omega'}}\sum_{\ell m}\sum_{\ell' m'}\nn\\
 &\quad :[-i\omega c_{\mu(s);\omega,\ell,m}Y^{\mu(s)}_{A(s)}Y_{\ell,m}e^{-i\omega u}+\hc][c_{\mu'(s);\omega',\ell',m'}Y^{\mu'(s)}_{A'(s)}Y_{\ell',m'}e^{-i\omega' u}+\hc]:\nn\\
 &=-i\int d\Omega Q^{\mu(s)\mu'(s)}\int_0^\infty d\omega\sum_{\ell m}\sum_{\ell' m'}c^\dagger_{\mu'(s);\omega,\ell',m'}c_{\mu(s);\omega,\ell,m}Y_{\ell,m}Y^*_{\ell',m'},\label{og1}
\end{align}
where the tensor $Q^{\mu(s)\mu'(s)}$ is the Cartesian version of $Q^{A(s)A'(s)}$
\bea Q^{\mu(s)\mu'(s)}=Q^{A(s)A'(s)}Y_{A(s)}^{\mu(s)}Y_{A'(s)}^{\mu'(s)}. 
\eea Equivalently, it is constructed from 
\begin{align}
 \gamma_{\mu\nu}=Y_\mu^AY_\nu^B\gamma_{AB}\quad \text{and}\quad \bar\epsilon_{\mu\nu}=Y^A_\mu Y^B_\nu \epsilon_{AB}.
\end{align}
The next step is using the bulk creation and annihilation operators to express the boundary ones (seeing \eqref{cb}), and we obtain
\begin{align}
 \co_{g=1} &=-i\int_0^\infty \frac{d^3\bm k}{(2\pi)^3} Q^{\mu(s)\mu'(s)}(\Omega_k)\sum_{\alpha\alpha'}\varepsilon_{\mu(s)}^{*\alpha}(\bm k)\varepsilon_{\mu'(s)}^{\alpha'}(\bm k)b^\dagger_{\alpha',\bm k}b_{\alpha,\bm k}.
\end{align}
We work in a representation where the particles have either right-hand or left-hand helicity. What follows is 
\begin{align}
 Q^{\mu(s)\mu'(s)}(\Omega_k)\varepsilon_{\mu(s)}^{*\alpha}(\bm k)\varepsilon_{\mu'(s)}^{\alpha'}(\bm k)=i\sigma_3^{\alpha\alpha'},\quad \alpha={\rm R,\, L},\label{eq628}
\end{align}
where $\sigma_3$ is the third Pauli matrix. Therefore, we find
\begin{align}
 \co_{g=1}&=\int\frac{d^3\bm k}{(2\pi)^3}(b^\dagger_{{\rm R},\bm k}b_{{\rm R},\bm k}-b^\dagger_{{\rm L},\bm k}b_{{\rm L},\bm k})\nn\\
 &=\int\frac{d^3\bm k}{(2\pi)^3}(n_{{\rm R},\bm k}-n_{{\rm L},\bm k}),
\end{align}
where $n_{\rm R/\rm L,\bm k}=b^\dagger_{{\rm R/\rm L},\bm k}b_{{\rm R/\rm L},\bm k}$ is the particle number with right/left-hand helicity. Therefore, 
 $\co_{g=1}$ is the difference between the numbers of particles with right-hand and left-hand helicity.

\section{Discussion and conclusion}\label{conc}
In this paper, we have reduced the bosonic Fronsdal theory in Minkowski spacetime to future null infinity $\mathcal{I}^+$. The boundary HS theory is characterized by the fundamental field $F_{A(s)}$ with a non-trivial symplectic form. All the descendants are determined by the fundamental field by the boundary constraint equations up to initial data. This extends the lower spin Carrollian field theories to general spin $s$. The symmetry algebra \eqref{trun}, which is formed by extending Poincar\'e and helicity flux operators, shows the same structure as the ones in the lower spin theories. All the flux operators are quadratic in the fundamental fields and could be interpreted as generators of supertranslation, superrotation and super-duality transformation, respectively. The super-duality transformation is the angle-dependent extension of the HS duality transformation \eqref{hsduality} at the null boundary. In Table \ref{corres}, we list the correspondences between the bulk global symmetry transformations and the boundary local transformations. 
\begin{table}
\begin{center}
\renewcommand\arraystretch{1.5}
 \begin{tabular}{|c||c|}\hline
Bulk global transformations &Boundary local transformations\\\hline\hline
Translation&Supertranslation\\\hline
Lorentz rotation&Superrotation\\\hline
Duality transformation&Super-duality transformation\\\hline
\end{tabular}
\caption{\centering{Bulk global transformations are extended to boundary local transformations}}\label{corres}
\end{center}
\end{table}
These results lead us to the conjecture that each bulk global symmetry transformation may extend to a boundary local symmetry transformation at the null hypersurfaces. These local symmetry transformations are related to the radiative flux operators from bulk to boundary. It would be interesting to check this conjecture in the future.
There are still many open questions to explore.
\begin{itemize}
 \item Further extension of the Carrollian diffeomorphism. There are HS extensions of BMS symmetry in the literature \cite{Campoleoni:2017mbt,Campoleoni:2020ejn,Campoleoni:2021blr,Bekaert:2022ipg} where the supertranslation and superrotation are large HS gauge transformations. The HS BMS algebra has been extended further for Carrollian conformal scalar theory \cite{Bekaert:2022oeh} which is expected to be dual to a non-trivial interacting HS theory in the bulk \cite{Boulanger:2023prx}. On the other hand, we work out the quantum flux operators following from Carrollian diffeomorphism which relates to spacetime geometry and differs from the ones concerning HS gauge fields. It would be interesting to see whether it is consistent to combine HS supertranslation and superrotation with Carrollian diffeomorphism. 
 \item General null hypersurfaces. The symmetry algebra found in this work should be valid for general null hypersurface, as has been shown in \cite{Li:2023xrr} for scalar theory. The general null hypersurface is intriguing since one may consider massive or non-flat spacetime HS theories. 
 \item Super-duality transformation. As has been mentioned in the introduction, duality transformations are found in various gravitational and gauge theories. It would be better to discuss their associated super-duality transformations on null boundaries. Besides, it is rather interesting to discuss the physical origin of super-duality transformation and its various consequences. 
\end{itemize}
\vspace{10pt}
{\noindent \bf Acknowledgments.} 
The work of J.L. is supported by NSFC Grant No. 12005069.
\appendix
\section{Number of independent components}\label{dof}
\begin{figure}
 \centering
 \begin{tikzpicture}
 \draw[](0,0)--(5,0)--(5,1)--(0,1)--(0,0);
 \draw[](1,0)--(1,1);\draw[](2,0)--(2,1);\draw[](3,0)--(3,1);\draw[](4,0)--(4,1);
 \node at (2.5,0.5){$\cdots$};
 \node at (5.5,0.5){$s$};
 \end{tikzpicture}
 \caption{\centering{Young diagram for a rank $s$ symmetric tensor}}
 \label{young}
\end{figure}
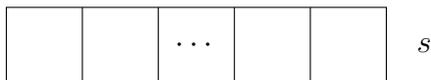

In this appendix, we will review the number of independent components for a symmetric tensor in $d$ dimensions. The results can be found in any book on the representation of Lie groups and we use the review reference \cite{Bekaert:2006py}. For a $d$-dimensional vector space $V$, the symmetric tensors of rank $s$ form a vector space $\text{Sym}^sV$. The number of independent degrees of freedom is equal to the dimension of the space $\text{dim}(\text{Sym}^sV)$. The symmetric tensor forms an irreducible representation of the general linear group $GL(d,\mathbb{R})$ and corresponds to the Young diagram with one row of length $s$ as shown in Figure \ref{young}.

\begin{figure}
 \centering
 \begin{tikzpicture}
 \draw[](0,0)--(6,0)--(6,1)--(0,1)--(0,-4)--(1,-4)--(1,1);\node at (6.5,0.5){$\lambda_1$};\node at (1.5,-1.5){$\cdots$};\node at (2.5,-2.5){$\vdots$};\node at (3.5,-1.5){$\cdots$};
 \node at (2.5,-1.5){$(i,j)$};\node at (2.5,-0.5){\vdots};\node at (4.5,-1.5){$\lambda_i$};
 \node at (1.5,-3.5){$\lambda_r$};
 \draw[](5,1)--(5,-1)--(4,-1)--(4,-2)--(3,-2)--(3,-3)--(2,-3);\draw[](3,-1)--(3,-2)--(2,-2);
 \draw[](2,1)--(2,-3);\draw[](3,1)--(3,-1)--(0,-1);\draw[](4,1)--(4,-1)--(3,-1);\draw[](0,-3)--(2,-3);\draw[](0,-2)--(2,-2);\node at (0.5,-3.5){$\cdots$}; \node at (5.5,0.5){$\cdots$};
 \end{tikzpicture}
 \caption{\centering{Young diagram of type $\lambda=(\lambda_1,\lambda_2,\cdots,\lambda_r)$}}
 \label{youngij}
\end{figure}
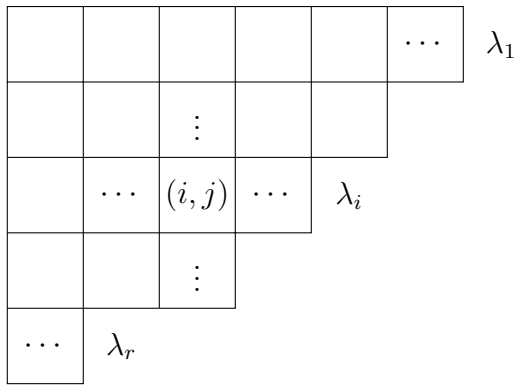

The dimension of any irreducible representation $V_\lambda$ of $GL(d,\mathbb{R})$ associated with Young diagram $\lambda=(\lambda_1,\lambda_2,\cdots,\lambda_r)$ is given by the formula 
\bea 
\text{dim}(V_\lambda)=\prod_{(i,j)}\frac{d-i+j}{\text{hook length}}
\eea 
where $(i,j)$ denotes the box in the $i$-th row and $j$-th column in the Young diagram, as shown in Figure \ref{youngij}. The product is over all boxes in the diagram and the hook length is the number of squares directly below or to the right of the square $(i,j)$, counting itself only once.
For the symmetric representation $\text{Sym}^sV$, there is only one row with $s$ boxes. Therefore, the dimension $\text{dim}(\text{Sym}^sV)$ is 
\bea 
\text{dim}(\text{Sym}^sV)=\frac{d(d+1)\cdots(d+s-1)}{s(s-1)\cdots 1}=C^s_{d+s-1}.
\eea

\section{Identities involving coordinate transformation }\label{identities}
In the context, we defined the four vectors in Minkowski spacetime as follows
\bea 
n^\mu=(1,n^i),\quad \bar{n}^\mu=(-1,n^i),\quad m^\mu=(0,n^i),\quad \bar{m}^\mu=(1,0)
\eea where $n^i$ is the normal vector of the unit sphere $S^2$. The vectors $Y_\mu^A$ are related to the first three vectors by 
\be 
Y_\mu^A=-\nabla^A n_\mu=-\nabla^A \bar{n}_\mu=-\nabla^A m_\mu.
\ee These vectors satisfy various identities which are collected in \cite{Li:2023xrr}. In this appendix, we can derive more 
identities associated with $N^\alpha_{\ \mu}$ and $\bar{N}^\mu_{\ \alpha}$ in the following
\begin{subequations}
\begin{align}
 Y^\mu_A N^\alpha_{\ \mu}&=-\delta^\alpha_A,\quad Y^A_\mu \bar{N}^\mu_{\ \alpha}=-\delta^A_\alpha,\\
 n^\mu N^\alpha_{\ \mu}&=\delta^\alpha_r,\quad n_\mu \bar N^\mu_{\ \alpha}=-\delta_\alpha^u,\\
 m^\mu N^\alpha_{\ \mu}&=\delta^\alpha_r-\delta^\alpha_u,\quad m_\mu \bar N^\mu_{\ \alpha}=\delta^r_\alpha,\\
 \bar{n}^\mu N^\alpha_{\ \mu}&=\delta^\alpha_r-2\delta^\alpha_u,\quad \bar{n}_\mu \bar N^\mu_{\ \alpha}=\delta^u_\alpha+2\delta^r_\alpha,\\
 \bar{m}^\mu N^\alpha_{\ \mu}&=\delta^\alpha_u,\quad \bar{m}_\mu \bar N^\mu_{\ \alpha}=-\delta^r_\alpha-\delta^u_\alpha,
\end{align}
\end{subequations}
and
\begin{subequations}
\begin{align}
 \nabla_AN^\alpha_{\ \mu}&=Y_{\mu A}(\delta^\alpha_u-\delta^\alpha_r)-m_\mu\delta^{\alpha}_A,\quad \nabla_A\bar N^\mu_{\ \alpha}=-Y^\mu_A\delta^r_\alpha-\gamma_{AB}m^\mu\delta_\alpha^B,\\
 Y^\mu_B\nabla^A N^\alpha_{\ \mu}&=\delta^A_B(\delta^\alpha_u-\delta^\alpha_r),\quad Y_\mu^B\nabla_A\bar{N}^\mu_{\ \alpha}=-\delta^B_A\delta^r_\alpha,\quad
 n^\mu\nabla_A N^\alpha_{\ \mu}=-\delta^\alpha_A,
\end{align}
\end{subequations}
as well as
\bs\begin{align}
 \bar{J}^\mu_{\ \alpha}N^\beta_{\ \mu}&=\delta^u_\alpha\delta^\beta_u+\delta^r_\alpha\delta^\beta_r+r\delta^A_\alpha \delta_A^\beta=\delta^\beta_\alpha+(r-1)\delta^A_\alpha \delta^\beta_A=r\delta^\beta_\alpha+(1-r)[\delta^u_\alpha \delta^\beta_u+\delta^r_\alpha\delta^\beta_r],\\
 N^\alpha_{\ \mu} \bar{N}^\mu_{\ \beta}&=\delta^\alpha_\beta,\quad \bar{N}^\mu_{\ \alpha}N^\alpha_{\ \nu}=\delta^\mu_\nu,\quad N^\alpha_{\ \mu} N^{\beta\mu}=-\delta^\alpha_u\delta^\beta_r-\delta^\beta_u\delta^\alpha_r+\delta^\alpha_r\delta^\beta_r+\gamma^{AB}\delta^\alpha_A\delta^\beta_B,\\
 \bar{N}^\mu_{\ \beta}\nabla_A N^\alpha_{\ \mu}&={ -}\delta^r_\beta \delta^\alpha_A+\gamma_{AB}\delta^B_\beta(\delta^\alpha_r-\delta^\alpha_u),{\quad N^\alpha_{\ \mu} \nabla_A N^{\beta\mu}=\delta^\alpha_A(\delta^\beta_r-\delta^\beta_u)-\delta^\beta_A(\delta^\alpha_r-\delta^\alpha_u)}.
\end{align}\es



\section{Asymptotic expansion of stress tensor near \texorpdfstring{$\mathcal{I}^+$}{}}\label{appc}
The partial derivatives of the HS gauge field are 
\bea 
\partial_\nu f_{\mu(s)}&&=\sum_{k=1}^\infty r^{-k}[-n_\nu N^{\alpha_1}_{\mu_1}\cdots N^{\alpha_s}_{\mu_s}\dot{F}^{(k)}_{\alpha(s)}-(k-1)m_\nu N^{\alpha_1}_{\mu_1}\cdots N^{\alpha_s}_{\mu_s}{F}^{(k-1)}_{\alpha(s)}-Y^A_{\nu}\nabla_A(N^{\alpha_1}_{\mu_1}\cdots N^{\alpha_s}_{\mu_s}{F}^{(k-1)}_{\alpha(s)})].\nn\\
\eea Therefore, we find the following quadratic terms consisting of the stress tensor
\bs\begin{align}
 \partial_\nu f_{\mu(s)}\partial^\nu f^{\mu(s)}&=\frac{2\dot{F}_{A(s)}F^{A(s)}}{r^3}+\cdots,\\
 \partial_\rho f_{\mu(s)}\partial_\sigma f^{\mu(s)}&=\frac{n_\rho n_\sigma\dot{F}_{A(s)}\dot{F}^{A(s)}}{r^2}+\frac{1}{r^3}[2n_\rho n_\sigma \dot{F}_{A(s)}\dot{F}^{(2)A(s)}+2n_{(\rho} m_{\sigma)}\dot{F}_{A(s)}F^{A(s)}\nn\\&+2n_{(\rho} Y_{\sigma)}^B\dot{F}_{A(s)}\nabla_B F^{A(s)}+2sn_{(\rho} Y_{\sigma)}^B(\dot{F}_{uA(s-1)}F_B^{\ A(s-1)}-\dot{F}_{BA(s-1)}F_u^{\ A(s-1)})]+\cdots ,\\
 \partial_\nu f_{\rho\mu(s-1)}\partial^\nu f_{\sigma}^{\ \mu(s-1)}&=\frac{1}{r^3}[2n_\rho n_\sigma \dot{F}_{uA(s-1)}F_u^{\ A(s-1)}+2n_{(\rho} Y_{\sigma)}^C\frac{d}{du}(F_{uA(s-1)}F_C^{\ A(s-1)})\nn\\&+Y_\rho^BY_\sigma^C\frac{d}{du}(F_{BA(s-1)}F_{C}^{\ A(s-1)})]+\cdots,\\
 \partial_\rho f_{\sigma\mu(s-1)}\partial^\sigma f^{\rho \mu(s-1)}&=\frac{2\dot{F}_{A(s)}F^{A(s)}}{r^3}+\cdots,\\
 \partial_\rho f^\nu_{\ \mu(s-1)}\partial_\nu f_\sigma^{\ \mu(s-1)}&=\frac{1}{r^3}[n_\rho n_\sigma (s \dot{F}_{uA(s-1)}F_u^{\ A(s-1)}{ +}\dot{F}_{A(s)}\nabla^{A_1}F_u^{\ A_2\cdots A_s})\nn\\&+n_\rho Y_\sigma^{B}(s\dot{F}_{uA(s-1)}F_{B}^{\ A(s-1)}-F_{uA(s-1)}\dot{F}_{B}^{\ A(s-1)}+\dot{F}_{CA(s-1)}\nabla^C F_{B}^{\ A(s-1)})\nn\\&+n_\sigma Y_\rho^{B}F_{BA(s-1)}\dot{F}_{u}^{\ A(s-1)}+n_\rho m_\sigma \dot{F}_{A(s)}F^{A(s)}+Y_\rho^{B}Y_\sigma^{C}\dot{F}_{CA(s-1)}F_{B}^{\ A(s-1)}]+\cdots,\\
 \partial_{\lambda} f^\nu_{\ \rho\mu(s-2)}\partial_\nu f^{\lambda\hspace{4pt}\mu(s-2)}_{\ \sigma}&=\frac{1}{r^3}\frac{d}{du}[N^\alpha_{\ \rho} F_{\alpha A(s-1)}N^\beta_{\ \sigma}F_{\beta}^{\ A(s-1)}]+\cdots.
\end{align}\es 
\section{Doubly symmetric traceless tensor on $S^2$}\label{st2}
We will study the doubly symmetric traceless tensors $X_{A(s)B(s)},Q_{A(s)B(s)}$ and $P_{AB(s)CD(s)}$ in this appendix. 

The trace-free representation of the fully symmetric, rank $k$ tensor $T_0^{a(k)}$ is given by the formula \cite{1965lgr..book.....T, Thorne:1980ru} in three dimensions and \cite{Toth:2021cpx} in general dimensions
\bea 
T^{a(k)}=T_0^{a(k)}+\sum_{p=1}^{[k/2]}(-1)^{p}\frac{k![d+2k-2(p+2)]!!}{2^{p}p!(k-2p)!(d+2k-4)!!}\eta^{(a_{1}a_{2}}\cdots \eta^{a_{2p-1}a_{2p}}T_p^{a_{2p+1}\cdots a_{k})}
\eea 
where $T_p^{a(k-2p)}$ is obtained by taking the trace of $T_0^{a(k)}$ $p$ times 
\bea 
T_p^{a_{2p+1}\cdots a_{k}}=\eta_{a_1a_2}\eta_{a_3a_4}\cdots\eta_{a_{2p-1}a_{2p}}T_0^{a_1a_2\cdots a_k}
\eea 
and $\eta^{a_1a_2}$ is the metric of the manifold. Note that the formula can be simplified to 
\bea 
T^{a(k)}=\sum_{p=0}^{[k/2]}(-1)^{p}\frac{k![d+2k-2(p+2)]!!}{2^{p}p!(k-2p)!(d+2k-4)!!}\eta^{(a_{1}a_{2}}\cdots \eta^{a_{2p-1}a_{2p}}T_p^{a_{2p+1}\cdots a_{k})}.
\eea In our case, i.e., $d=2, k=s$, the trace-free part of a fully symmetric, rank $s$ tensor $T_0^{A(s)}$ is 
\bea 
T^{A(s)}=\sum_{p=0}^{[s/2]}a(p;s)\gamma^{(A_{1}A_{2}}\cdots \gamma^{A_{2p-1}A_{2p}}T_p^{A_{2p+1}\cdots A_{s})}
\eea where 
\bea 
a(p;s)=(-1)^{p}\frac{s![2s-2p-2]!!}{2^{p}p!(s-2p)!(2s-2)!!},\quad p=0,1,\cdots,[s/2].
\eea For later convenience, we extend the definition of $a(p;s)$ to $p=-1$ with 
\be 
a(-1;s)=0.
\ee In \cite{Toth:2021cpx}, this is checked up to rank 8 by computer. 
It may be proved by noticing the identity 
\bea 
&&\gamma_{A_1A_2}\gamma^{(A_{1}A_{2}}\cdots \gamma^{A_{2p-1}A_{2p}}T_p^{A_{2p+1}\cdots A_{s})}\nn\\&&=b(p;s)\gamma^{(A_3A_4}\cdots\gamma^{A_{2p-1}A_{2p}}T_p^{A_{2p+1}\cdots A_s)}+c(p+1;s)\gamma^{(A_3A_4}\cdots\gamma^{A_{2p+1}A_{2p+2}}T_{p+1}^{A_{2p+3}\cdots A_s)}
\eea 
with 
\bs 
\begin{align}
 b(p;s)&=\frac{4p(s-p)}{s(s-1)},\\
 c(p+1;s)&=\frac{(s-2p)(s-2p-1)}{s(s-1)}. 
\end{align}
\es
The coefficients $a(p;s),b(p;s)$ and $c(p;s)$ satisfy the identity 
\be 
a(p;s)b(p;s)+a(p-1;s)c(p;s)=0,\quad p=0,1,\cdots, [s/2].
\ee 
Note that we have used $a(-1;s)=0$ in the above equation. Therefore, the trace vanishes
\bea 
\gamma_{A_1A_2}T^{A_1\cdots A_s}&&=\sum_{p=0}^{[s/2]}a(p;s)[b(p;s)\gamma^{(A_3A_4}\cdots\gamma^{A_{2p-1}A_{2p}}T_p^{A_{2p+1}\cdots A_s)}\nn\\&&\hspace{1.7cm}+c(p+1;s)\gamma^{(A_3A_4}\cdots\gamma^{A_{2p+1}A_{2p+2}}T_{p+1}^{A_{2p+3}\cdots A_s)}]\nn\\&&=\sum_{p=0}^{[s/2]}[a(p;s)b(p;s)+a(p-1;s)c(p;s)]\gamma^{(A_3A_4}\cdots\gamma^{A_{2p-1}A_{2p}}T_p^{A_{2p+1}\cdots A_s)}\nn\\&&=0.
\eea 
\subsection{Doubly symmetric tensor \texorpdfstring{$X_{A(s)B(s)}$}{}}
Now we will prove the formula \eqref{Xsym} in the context. 
 Introducing the notation 
\bea 
\widetilde{X}_{p,q}^{A_{2p+1}\cdots A_sB_{2q+1}\cdots B_s}=\gamma_{A_1A_2}\cdots \gamma_{A_{2p-1}A_{2p}}\gamma_{B_1B_2}\cdots\gamma_{B_{2q-1}B_{2q}}\widetilde{X}^{(A_1\cdots A_s)(B_1\cdots B_s)},
\eea this is obtained by taking traces $p$ and $q$ times for the indices $A$s and $B$s, respectively. When $p=q=0$, we have 
\be 
\widetilde{X}^{A_1\cdots A_sB_1\cdots B_s}_{0,0}=\widetilde{X}^{(A_1\cdots A_s)(B_1\cdots B_s)}.
\ee 
We use the vielbeins $e^A_{\hat a}$ to decompose the metric $\gamma^{AB}$ as 
\be 
\gamma^{AB}=e^A_{\hat a} e^{B \hat a}\label{veinbein}
\ee and thus 
\bea 
\gamma^{A_1B_1}\cdots\gamma^{A_sB_s}=e^{A_1}_{\hat a_1}\cdots e^{A_s}_{\hat a_s}e^{B_1\hat{a}_1}\cdots e^{B_s\hat a_s}.
\eea It follows that 
\bea 
\widetilde{X}^{(A_1\cdots A_s)(B_1\cdots B_s)}&&=\frac{1}{s!}\sum_{\pi\in S_s}\gamma^{A_{\pi(1)}B_1}\cdots\gamma^{A_{\pi(s)}B_s}\nn\\&&=\frac{1}{s!}\sum_{\pi\in S_s}e^{A_{\pi(1)}}_{\hat a_1}\cdots e^{A_{\pi(s)}}_{\hat a_s}e^{B_1\hat a_1}\cdots e^{B_s\hat a_s}\nn\\&&=V^{A_1\cdots A_s}_{\hat{a}_1\cdots \hat{a}_s} e^{B_1\hat a_1}\cdots e^{B_s\hat a_s}.
\eea In the first step, the indices $A_1\cdots A_s$ are symmetrized using the permutation group $S_s$. In the second step, we use the formula \eqref{veinbein}. In the last step, we define the symmetric tensor 
\be 
V^{A_1\cdots A_s}_{\hat a_1\cdots \hat a_s}=\frac{1}{s!}\sum_{\pi\in S_s}e^{A_{\pi(1)}}_{\hat a_1}\cdots e^{A_{\pi(s)}}_{\hat a_s}
\ee where
\be 
V^{A_1\cdots A_s}_{\hat a_1\cdots\hat a_s}=V^{(A_1\cdots A_s)}_{\hat a_1\cdots\hat a_s}=V^{A_1\cdots A_s}_{(\hat a_1\cdots\hat a_s)}.
\ee Therefore, the indices $B_1\cdots B_s$ is symmetrized automatically. We may rewrite 
\be 
\widetilde{X}^{(A_1\cdots A_s)(B_1\cdots B_s)}=V^{A_1\cdots A_s}_{\hat a_1\cdots \hat a_s}V^{B_1\cdots B_s\hat a_1\cdots\hat a_s}.
\ee The $p$-th trace of $V^{A_1\cdots A_s}_{a_1\cdots a_s}$ is 
\bea V_{p;\hat a_1\cdots a_s}^{A_{2p+1}\cdots A_s}\equiv 
\gamma_{A_1A_2}\cdots\gamma_{A_{2p-1}A_{2p}}V^{A_1\cdots A_s}_{\hat a_1\cdots \hat a_s}.
\eea We could find the following product
\bea 
 V_{p;\hat a_1\cdots a_s}^{A_{2p+1}\cdots A_s}V_{q}^{B_{2q+1}\cdots B_s\hat a_1\cdots\hat a_s}&&=\gamma_{A_1A_2}\cdots\gamma_{A_{2p-1}A_{2p}}V^{A_1\cdots A_s}_{\hat a_1\cdots \hat a_s}\gamma_{B_1B_2}\cdots\gamma_{B_{2q-1}B_{2q}}V^{B_1\cdots B_s \hat a_1\cdots\hat a_s}\nn\\&&=\widetilde{X}_{p,q}^{A_{2p+1}\cdots A_s B_{2q+1}\cdots B_{s}}.
\eea 
Then the trace-free part of the tensor $\widetilde{X}^{(A_1\cdots A_s)(B_1\cdots B_s)}$ with respect to $A(s)$ and $B(s)$ is 
\bea 
X^{A(s)B(s)}&&=\sum_{p=0}^{[s/2]}a(p;s)\gamma^{(A_{1}A_{2}}\cdots \gamma^{A_{2p-1}A_{2p}}V_{p;\hat a_1\cdots \hat a_s}^{A_{2p+1}\cdots A_s)} \sum_{q=0}^{[s/2]}a(q;s)\gamma^{(B_1B_2}\cdots\gamma^{ B_{2q-1}B_{2q}}V^{B_{2q+1}\cdots B_s)\hat a_1\cdots \hat a_s}_q\nn\\&&=\sum_{p,q=0}^{[s/2]}a(p;s)a(q;s)\gamma^{(A_{1}A_{2}}\cdots \gamma^{A_{2p-1}A_{2p}}V_{p;\hat a_1\cdots \hat a_s}^{A_{2p+1}\cdots A_s)}\gamma^{(B_1B_2}\cdots\gamma^{ B_{2q-1}B_{2q}}V^{B_{2q+1}\cdots B_s)\hat a_1\cdots \hat a_s}_q\nn\\&&=\sum_{p,q=0}^{[s/2]}a(p,q;s)\gamma^{(A_{1}A_{2}}\cdots \gamma^{A_{2p-1}A_{2p}}\widetilde{X}_{p,q}^{A_{2p+1}\cdots A_{s})(B_{2q+1}\cdots B_s}\gamma^{B_1B_2}\cdots\gamma^{B_{2q-1}B_{2q})}\label{stformula}
\eea 
with the coefficients $a(p,q;s)$ being 
\be 
a(p,q;s)=a(p;s)a(q;s)=(-1)^{p+q}\frac{s![2s-2p-2]!!}{2^{p}p!(s-2p)!(2s-2)!!}\frac{s![2s-2q-2]!!}{2^{q}q!(s-2q)!(2s-2)!!}.
\ee 

As a consistency check, we will list the tensors $X^{A(s)B(s)}$ for several cases of lower spin in the following.
\begin{enumerate}
 \item $s=2$. The tensor $\widetilde{X}^{(A_1A_2)(B_1B_2)}$ is
\be 
\widetilde{X}^{A_1A_2B_1B_2}=\frac{1}{2}(\gamma^{A_1B_1}\gamma^{A_2B_2}+\gamma^{A_1B_2}\gamma^{A_2B_1})
\ee whose traces are 
\bea 
\widetilde{X}_{1,0}^{B_1B_2}=\gamma^{B_1B_2},\quad \widetilde{X}_{0,1}^{A_1A_2}=\gamma^{A_1A_2},\quad \widetilde{X}_{1,1}=2.
\eea With the formula \eqref{stformula}, we find 
\bea 
X^{A_1A_2B_1B_2}=\frac{1}{2}(\gamma^{A_1B_1}\gamma^{A_2B_2}+\gamma^{A_1B_2}\gamma^{A_2B_1})-\frac{1}{2}\gamma^{A_1A_2}\gamma^{B_1B_2}.
\eea 

\item $s=3$. The tensor $\widetilde{X}^{(A_1A_2A_3)(B_1B_2B_3)}$ is
\bea 
\widetilde X^{(A_1A_2A_3)(B_1B_2B_3)}&=\frac{1}{3!}(\gamma^{A_1B_1}\gamma^{A_2B_2}\gamma^{A_3B_3}+\gamma^{A_1B_1}\gamma^{A_3B_2}\gamma^{A_2B_3}+\gamma^{A_2B_1}\gamma^{A_1B_2}\gamma^{A_3B_3}\nn\\
 &+\gamma^{A_2B_1}\gamma^{A_3B_2}\gamma^{A_1B_3}+\gamma^{A_3B_1}\gamma^{A_2B_2}\gamma^{A_1B_3}+\gamma^{A_3B_1}\gamma^{A_1B_2}\gamma^{A_2B_3}),
\eea and its various traces are
\bs\begin{align}
 \widetilde X_{1,0}^{A_iB_1B_2B_3}&=\gamma^{A_i(B_1}\gamma^{B_2B_3)},\quad \widetilde{X}_{0,1}^{(A_1A_2A_3)B_i}=\gamma^{(A_1A_2}\gamma^{A_3)B_i},\quad i=1,2,3,\\
 \widetilde{X}_{1,1}^{A_iB_j}&=\frac{4}{3}\gamma^{A_iB_j},\quad i,j=1,2,3.
\end{align}\es Therefore, the doubly symmetric traceless tensor $X^{A_1A_2A_3B_1B_2B_3}$ should be 
\begin{align}
 X^{A_1A_2A_3B_1B_2B_3}&=\widetilde{X}^{(A_1A_2A_3)(B_1B_2B_3)}-\frac{3}{4}\gamma^{(A_1A_2}\widetilde{X}^{A_3)(B_1B_2B_3)}_{1,0}-\frac{3}{4}\gamma^{(B_1B_2}\widetilde{X}^{B_3)(A_1A_2A_3)}_{0,1}\nn\\
&+\frac{9}{16}\gamma^{(A_1A_2}\widetilde{X}^{A_3)(B_1}_{1,1}\gamma^{B_2B_3)}.
\end{align}

\item $s=4$, the symmetric tensor $\widetilde{X}^{(A_1\cdots A_4)(B_1\cdots B_4)}$ is 
\bea 
\widetilde{X}^{(A_1\cdots A_4)(B_1\cdots B_4)}&=\frac{1}{4!}(\gamma^{A_1B_1}\gamma^{A_2B_2}\gamma^{A_3B_3}\gamma^{A_4B_4}+\text{permutations of}\ A_1A_2A_3A_4)
\eea and its various traces are 
\bs\begin{align}
 \widetilde{X}^{A_iA_jB_1B_2B_3B_4}_{1,0}&=\frac{1}{12}[\gamma^{B_1B_2}(\gamma^{A_iB_3}\gamma^{A_jB_4}+\gamma^{A_iB_4}\gamma^{A_jB_3})+\gamma^{B_1B_3}(\gamma^{A_iB_2}\gamma^{A_jB_4}+\gamma^{A_iB_4}\gamma^{A_jB_2})\nn\\&+\gamma^{B_1B_4}(\gamma^{A_iB_2}\gamma^{A_jB_3}+\gamma^{A_iB_3}\gamma^{A_jB_2})+\gamma^{B_2B_3}(\gamma^{A_iB_1}\gamma^{A_jB_4}+\gamma^{A_iB_4}\gamma^{A_jB_1})\nn\\&+\gamma^{B_2B_4}(\gamma^{A_iB_1}\gamma^{A_jB_3}+\gamma^{A_iB_3}\gamma^{A_jB_1})+\gamma^{B_3B_4}(\gamma^{A_iB_1}\gamma^{A_jB_2}+\gamma^{A_iB_2}\gamma^{A_jB_1})],\\
 \widetilde{X}^{B_1B_2B_3B_4}_{2,0}&=\frac{1}{3}(\gamma^{B_1B_2}\gamma^{B_3B_4}+\gamma^{B_1B_3}\gamma^{B_2B_4}+\gamma^{B_1B_4}\gamma^{B_2B_3}),\\
 \widetilde{X}^{A_1A_2A_3A_4B_iB_j}_{0,1}&=\frac{1}{12}[\gamma^{A_1A_2}(\gamma^{B_iA_3}\gamma^{B_jA_4}+\gamma^{B_iA_4}\gamma^{B_jA_3})+\gamma^{A_1A_3}(\gamma^{B_iA_2}\gamma^{B_jA_4}+\gamma^{B_iA_4}\gamma^{B_jA_2})\nn\\&+\gamma^{A_1A_4}(\gamma^{B_iA_2}\gamma^{B_jA_3}+\gamma^{B_iA_3}\gamma^{B_jA_2})+\gamma^{A_2A_3}(\gamma^{B_iA_1}\gamma^{B_jA_4}+\gamma^{B_iA_4}\gamma^{B_jA_1})\nn\\&+\gamma^{A_2A_4}(\gamma^{B_iA_1}\gamma^{B_jA_3}+\gamma^{B_iA_3}\gamma^{B_jA_1})+\gamma^{A_3A_4}(\gamma^{B_iA_1}\gamma^{B_jA_2}+\gamma^{B_iA_2}\gamma^{B_jA_1})],\\
 \widetilde{X}^{A_1A_2A_3A_4}_{0,2}&=\frac{1}{3}(\gamma^{A_1A_2}\gamma^{A_3A_4}+\gamma^{A_1A_3}\gamma^{A_2A_4}+\gamma^{A_1A_4}\gamma^{A_2A_3}),\\
 \widetilde{X}^{A_iA_jB_kB_l}_{1,1}&=\frac{1}{6}[\gamma^{A_iA_j}\gamma^{B_kB_l}+3(\gamma^{A_iB_k}\gamma^{A_jB_l}+\gamma^{A_iB_l}\gamma^{A_jB_k})],\\
 \widetilde{X}^{A_iA_j}_{1,2}&=\frac{4}{3}\gamma^{A_iA_j},\\
 \widetilde{X}^{B_iB_j}_{2,1}&=\frac{4}{3}\gamma^{B_iB_j},\\
 \widetilde{X}_{2,2}&=\frac{8}{3}.
\end{align}\es The doubly symmetric traceless tensor $X^{A_1A_2A_3A_4B_1B_2B_3B_4}$ should be
\bea 
&&X^{A_1A_2A_3A_4B_1B_2B_3B_4}\nn\\&&=\widetilde{X}^{(A_1A_2A_3A_4)(B_1B_2B_3B_4)}-\gamma^{(A_1A_2}\widetilde{X}_{1,0}^{A_3A_4)B_1B_2B_3B_4}-\widetilde{X}_{0,1}^{A_1A_2A_3A_4(B_3B_4}\gamma^{B_1B_2)}\nn\\&&+\gamma^{(A_1A_2}\widetilde{X}_{1,1}^{A_3A_4)(B_1B_2}\gamma^{B_3B_4)}+\frac{1}{8}\gamma^{(A_1A_2}\gamma^{A_3A_4)}\widetilde{X}_{2,0}^{B_1B_2B_3B_4}+\frac{1}{8}\gamma^{(B_1B_2}\gamma^{B_3B_4)}\widetilde{X}_{0,2}^{A_1A_2A_3A_4}\nn\\&&-\frac{1}{8}\gamma^{(A_1A_2}\gamma^{A_3A_4)}\gamma^{(B_1B_2}\widetilde{X}_{2,1}^{B_3B_4)}-\frac{1}{8}\gamma^{(B_1B_2}\gamma^{B_3B_4)}\gamma^{(A_1A_2}\widetilde{X}_{1,2}^{A_3A_4)}\nn\\
&&+\frac{1}{24}\gamma^{(A_1A_2}\gamma^{A_3A_4)}\gamma^{(B_1B_2}\gamma^{B_3B_4)}.\nn\\
\eea 
\end{enumerate}

\subsection{Other doubly symmetric traceless tensors and related identities}
In \eqref{doubleystP}, we also defined a doubly symmetric traceless tensor $P_{AB(s)CD(s)}$. Using the identity 
\bea
\gamma_{AC}\gamma_{BD}+s\gamma_{AB}\gamma_{CD}-s\gamma_{AD}\gamma_{CB}=\gamma_{AC}\gamma_{BD}+s\epsilon_{AC}\epsilon_{BD},
\eea we find
\bea 
\widetilde{P}_{AB_1\cdots B_sCD_1\cdots D_s}=(\gamma_{AC}\gamma_{B_1D_1}+s\epsilon_{AC}\epsilon_{B_1D_1})\gamma_{B_2D_2}\cdots \gamma_{B_sD_s},
\eea 
and the doubly symmetric traceless tensor becomes
\bea 
{P}_{AB(s)CD(s)}=\gamma_{AC}X_{B(s)D(s)}-s\epsilon_{AC}Q_{B(s)D(s)}.
\eea 
There are various identities associated with the tensor $X_{A(s)B(s)}, Q_{A(s)B(s)}$ and $P_{AB(s)CD(s)}$. 
\begin{enumerate}
 \item To calculate the commutator between $\mathcal{T}_f$ and the fundamental field $F_{A(s)}$, we need the identity 
 \bea 
 X_{A(s)B(s)}F^{B(s)}=F_{A(s)}.
 \eea 
 \item To compute the commutator between $\mathcal{M}_Y$ and the fundamental field $F_{A(s)}$, we need the identity 
 \bea 
 P_{AB(s)CD(s)}X^{B(s)}_{\hspace{0.45cm}E(s)}=P_{AE(s)CD(s)}.
 \eea 
 \item To calculate the commutator between $\mathcal{O}_g$ and the fundamental field $F_{A(s)}$, we need the identities 
 \bs\begin{align}
 & Q^{B(s)}_{\hspace{0.45cm}C(s)}X_{B(s)A(s)}=Q_{A(s)C(s)},\\
 &Q_{B(s)A(s)}=-Q_{A(s)B(s)}.
 \end{align}\es 
 \item By exchanging the indices $A$ and $C$ in the tensor $P_{AB(s)CD(s)}$, we find the tensor $\rho_{AB(s)CD(s)}$ 
 \bea 
 \rho_{AB(s)CD(s)}=\frac{1}{2}(P_{AB(s)CD(s)}+P_{AD(s)CB(s)})=\gamma_{AC}X_{B(s)D(s)}.
 \eea 
 \item To obtain the commutator $[\mathcal{T}_f,\mathcal{M}_Y]$, we used the following identities 
 \bs\begin{align}
 & \Delta_{A(s)}(fY;F)=f\Delta_{A(s)}(Y;F)+\frac{1}{2}Y^D\nabla^C f F^{B(s)}P_{DB(s)CA(s)},\\
 & \Delta_{A(s)}(Y;fF)=f\Delta_{A(s)}(Y;F)+Y^D\nabla_D f F_{A(s)},\\
 & \rho_{AB(s)CD(s)}\dot{F}^{B(s)}\dot{F}^{D(s)}=\gamma_{AC}\dot{F}^{B(s)}\dot{F}_{B(s)},\\
 & P_{AB(s)CD(s)}\dot{F}^{D(s)}F^{B(s)}=\gamma_{AC}\dot{F}^{B(s)}F_{B(s)}-s\epsilon_{AC}\epsilon^{DE}\dot{F}_D^{\hspace{2pt}B(s-1)}F_{EB(s-1)}.
 \end{align}\es \item For the commutator $[\mathcal{M}_Y,\mathcal{O}_g]$, we need the integral identity 
 \bea 
 \int d\Omega Q_{A(s)B(s)}gF^{B(s)}\Delta^{A(s)}(Y;\dot F)=-\int d\Omega Q_{A(s)B(s)}\dot F^{A(s)}\Delta^{B(s)}(Y;g F)
 \eea which follows from the algebraic identity 
 \bea 
 Q_{E(s)A(s)}P_{DB(s)C}^{\hspace{1cm}A(s)}+Q_{A(s)B(s)}P_{DE(s)C}^{\hspace{1cm}A(s)}=2\gamma_{CD}Q_{E(s)B(s)}.
 \eea 
 \item For the commutator $[\mathcal{M}_Y,\mathcal{M}_Z]$, we need the identity 
 \bea 
 \Delta_{A(s)}(Y;\Delta(Z;F))-\Delta_{A(s)}(Z;\Delta(Y;F))-\Delta_{A(s)}([Y,Z];F)=s o(Y,Z)Q_{B(s)A(s)}F^{B(s)}.\nn\\\label{mymzo}
 \eea To prove this formula, we may rewrite the left-hand side as 
 \bea 
 \text{LHS}&&=\text{terms with $F$}+\text{terms with $\nabla F$}+\text{terms with $\nabla\nabla F$}.
 \eea 
 The terms with the second derivative of $F$ are 
 \bea 
 && Z^GY^D\nabla^C\nabla^H F^{E(s)} \rho_{GE(s)H}^{\hspace{1cm}B(s)}\rho_{DB(s)CA(s)}-(Y\leftrightarrow Z)\nn\\&&=Z^DY^C[\nabla_C,\nabla_D]F_{A(s)}\nn\\&&=-Z^DY^C R^E_{\ A_1CD}F_{EA_2\cdots A_s}-\cdots -Z^DY^C R^E_{\ A_sCD}F_{A_1\cdots A_{s-1}E}
 \nn\\&&=-s Y^CZ^DR_{CDE(A_1}F_{A_2\cdots A_s)}^{\hspace{1cm}E}\label{nabla2}
 \eea 
 The terms with only the first derivative of $F$ are
 \bea 
 && [Y^D\nabla^CZ^G \nabla^H F^{E(s)}\rho_{GE(s)H}^{\hspace{0.9cm} \ B(s)}\rho_{DB(s)CA(s)}+\frac{1}{2}Y^D\nabla^HZ^G\nabla^C F^{E(s)} P_{GE(s)H}^{\hspace{0.9cm} \ B(s)}\rho_{DB(s)CA(s)}\nn\\&&+\frac{1}{2}Z^G\nabla^CY^D\nabla^H F^{E(s)} P_{DB(s)CA(s)}\rho_{GE(s)H}^{\hspace{0.9cm} \ B(s)}]-(Y\leftrightarrow Z)-[Y,Z]^D\nabla^C F^{B(s)}\rho_{DB(s)CA(s)}\nn\\&&=0.
 \eea 
 The terms linear in $F$ are 
 \bea 
 &&[\frac{1}{2}Y^C\nabla_C\nabla^H Z^G\ F^{E(s)}P_{GE(s)HA(s)}+\frac{1}{4}\nabla^CY^D \nabla^HZ^G F^{E(s)}P_{GE(s)H}^{\hspace{1cm}B(s)}P_{DB(s)CA(s)}]\nn\\&&-(Y\leftrightarrow Z)-\frac{1}{2}\nabla^C[Y,Z]^D F^{B(s)}P_{DB(s)CA(s)}\nn\\&&=\text{terms with $\nabla\nabla Y$ or $\nabla\nabla Z$}+\text{terms with $\nabla Y\ \nabla Z$}
 \eea 
 where the first part can be turned into commutators 
 \bea 
 && \text{terms with $\nabla\nabla Y$ or $\nabla\nabla Z$}\nn\\&&=\frac{1}{2}Y^C\nabla_C\nabla^H Z^G\ F^{B(s)} P_{G B(s)HA(s)}-\frac{1}{2}Z^C\nabla_C \nabla^H Y^G F^{B(s)}P_{GB(s)HA(s)}\nn\\&&-\frac{1}{2}Y^H\nabla^C\nabla_HZ^D F^{B(s)}P_{DB(s)HA(s)}+\frac{1}{2}Z^H\nabla^C\nabla_H Y^D F^{B(s)}P_{DB(s)CA(s)}\nn\\&&=\frac{1}{2}Y^C[\nabla_C,\nabla^H]Z^D F^{B(s)}P_{DB(s)HA(s)}-(Y\leftrightarrow Z)\nn\\&&=\frac{1}{2}Y^CZ^ER^{D\hspace{8pt}H}_{\ EC}F^{B(s)}P_{DB(s)HA(s)}-(Y\leftrightarrow Z)\nn\\&&=\frac{s}{2}Y^CZ^E(R_{DEC(A_1}+R_{CDE(A_1})F_{A_2\cdots A_s)}^{\hspace{1cm} D}-(Y\leftrightarrow Z).
 \eea Utilizing the Bianchi identity $R_{A[BCD]}=0$, we find that the above results are canceled by \eqref{nabla2}. 
 With the identity 
 \bea 
 P_{GE(s)H}^{\hspace{1cm}B(s)}P_{DB(s)CA(s)} = P_{DE(s)C}^{\hspace{1cm}B(s)}P_{GB(s)HA(s)},
 \eea the terms with $\nabla Y$ and $\nabla Z$ are 
 \bea 
 &&\text{ terms with $\nabla Y$ and $\nabla Z$}\nn\\&&=\frac{1}{2}\nabla^EY^F\nabla^GZ^I F^{B(s)}(\gamma_{IE}P_{FB(s)GA(s)}-\gamma_{FG}P_{IB(s)EA(s)})\nn\\&&=\frac{s}{2}\nabla^EY^F \nabla^GZ^I F^{B(s)}(\gamma_{FG}\epsilon_{IE}-\gamma_{IE}\epsilon_{FG})Q_{A(s)B(s)}\nn\\&&=-\frac{s}{4}\nabla^AY^B \nabla^CZ^D (\epsilon^{BC}\gamma^{AD}+\epsilon^{AC}\gamma^{BD}+\epsilon^{BD}\gamma^{AC}+\epsilon^{AD}\gamma^{BC})Q_{A(s)B(s)}F^{B(s)}\nn\\&&=s o(Y,Z)Q_{B(s)A(s)}F^{B(s)}.
 \eea We have used the Fierz identity 
 \be 
 \gamma_{AB}\epsilon_{CD}+\gamma_{AC}\epsilon_{DB}+\gamma_{AD}\epsilon_{BC}=0.
 \ee Therefore, we finish the proof of the identity \eqref{mymzo}.
 \item To compute the central charge in the commutator $[\mathcal{T}_{f_1},\mathcal{T}_{f_2}]$, we need the square of $X_{A(s)B(s)}$
 \be 
 X_{A(s)B(s)}X^{A(s)B(s)}=2.
 \ee 
 \item To compute the central charge ${\rm C}_{MO}(Y,g)$, we need the identity
 \bea 
 P_{AB(s)CD(s)}Q^{B(s)D(s)}=-2s\epsilon_{AC}
 \eea which follows from the identities
 \bea 
 X_{A(s)B(s)}Q^{A(s)B(s)}=0,\quad Q_{A(s)B(s)}Q^{A(s)B(s)}=2.
 \eea 
\end{enumerate}

\subsection{Tensors and identities in stereographic project coordinates}
The previous identities may be checked in stereographic project coordinates.
The non-vanishing components of the doubly symmetric traceless tensors in this coordinate system are 
\bs\begin{align}
 & X_{z(s)\bar{z}(s)}=X_{\bar{z}(s)z(s)}=\gamma^s,\\
 & P_{Az(s) C\bar{z}(s)} =\gamma^s(\gamma_{AC}-is\epsilon_{AC}),\\
 & P_{A\bar{z}(s) C z(s)} =\gamma^s(\gamma_{AC}+is\epsilon_{AC}),\\
 & Q_{z(s)\bar{z}(s)}=-Q_{\bar{z}(s)z(s)}=i\gamma^s,\\
 & \rho_{Az(s)C\bar{z}(s)}=\rho_{A\bar{z}(s)Cz(s)}=\gamma_{AC}\gamma^s
\end{align}\es 
For example, the square of $X_{A(s)B(s)}$ can be found to be 
\be 
X_{A(s)B(s)}X^{A(s)B(s)}=X_{z(s)\bar{z}(s)}X^{z(s)\bar{z}(s)}+X_{\bar{z}(s)z(s)}X^{\bar{z}(s)z(s)}=2.
\ee 
One can also use the coordinate transformation to find the doubly symmetric traceless tensors, e.g.,
\bs\begin{align}
 X_{A(s)B(s)}&=\frac{\partial z(s)}{\partial \theta^{A(s)}}\frac{\partial\bar{z}(s)}{\partial\theta^{B(s)}}X_{z(s)\bar{z}(s)}+\frac{\partial \bar z(s)}{\partial \theta^{A(s)}}\frac{\partial{z}(s)}{\partial\theta^{B(s)}}X_{\bar z(s){z}(s)}=\gamma^s\left(\frac{\partial z(s)}{\partial \theta^{A(s)}}\frac{\partial\bar{z}(s)}{\partial\theta^{B(s)}}+\frac{\partial \bar z(s)}{\partial \theta^{A(s)}}\frac{\partial{z}(s)}{\partial\theta^{B(s)}}\right),\\
 Q_{A(s)B(s)}&=\frac{\partial z(s)}{\partial \theta^{A(s)}}\frac{\partial\bar{z}(s)}{\partial\theta^{B(s)}}Q_{z(s)\bar{z}(s)}+\frac{\partial \bar z(s)}{\partial \theta^{A(s)}}\frac{\partial{z}(s)}{\partial\theta^{B(s)}}Q_{\bar z(s){z}(s)}=i\gamma^s\left(\frac{\partial z(s)}{\partial \theta^{A(s)}}\frac{\partial\bar{z}(s)}{\partial\theta^{B(s)}}-\frac{\partial \bar z(s)}{\partial \theta^{A(s)}}\frac{\partial{z}(s)}{\partial\theta^{B(s)}}\right).
\end{align}\es

In terms of the projective stereographic coordinates, the flux density operators are simplified greatly
\bs\begin{align}
 T(u,z,\bar z)&=2\gamma^{-s}\dot{F}\dot{\bar F},\\
 M_z(u,z,{\bar z})&={ \frac{1}{2}(1-s)\gamma^{-s}(\dot{\bar F}\nabla_{z}F-{\bar F}\nabla_{z}\dot F)+\frac{1}{2}(1+s)\gamma^{-s}(\dot{F}\nabla_{z}\bar F-{F}\nabla_{z}\dot{\bar F})},\\
 M_{\bar{z}}(u,z,\bar{z})&={ \frac{1}{2}(1+s)\gamma^{-s}(\dot{\bar F}\nabla_{\bz}F-{\bar F}\nabla_{\bz}\dot F)+\frac{1}{2}(1-s)\gamma^{-s}(\dot{F}\nabla_{\bz}\bar F-{F}\nabla_{\bz}\dot{\bar F})},\\ 
 O(u,z,\bar z)&={ i\gamma^{-s}(\dot{\bar F}F-\dot{F}\bar F)}.
\end{align}\es

\section{Canonical quantization}\label{canoquan}
\subsection{Mode expansion}
We can also use the mode expansion to quantize the fundamental field. After imposing the De Donder gauge, the EOM becomes a wave equation whose solution can be expanded in terms of plane waves
\begin{equation}
 f_{\mu(s)}(t,\bm x)=\sum_{\alpha}\int \frac{d^3\bm k}{(2\pi)^3}\frac{1}{\sqrt{2\omega_{\bm k}}}[\varepsilon^{*\alpha}_{\mu(s)}(\bm k)b_{\alpha,\bm k}e^{-i\omega t+i\bm k\cdot\bm x}+\varepsilon^{\alpha}_{\mu(s)}(\bm k)b^\dagger_{\alpha,\bm k}e^{i\omega t-i\bm k\cdot\bm x}], 
\end{equation}
where $\varepsilon^{\alpha}_{\mu(s)}(\bm k)$ is the polarization tensor.
Here the creation and annihilation operators satisfy the canonical commutator
\begin{equation}
 [b_{\alpha,\bm k}, b^\dagger_{\beta,\bm k'}]=(2\pi)^3\delta_{\alpha,\beta}\delta^{(3)}(\bm k-\bm k'), 
\end{equation}
while other commutators vanish. One can choose appropriate polarization tensors such that they obey the following completeness relation 
\begin{equation}
 \sum_{\alpha,\beta}\varepsilon_{\mu(s)}^{*\alpha}(\bm k)\delta_{\alpha,\beta}\varepsilon_{\nu(s)}^{\beta}(\bm k)=X_{\mu(s)\nu(s)},\label{comp}
\end{equation}
where $X_{\mu(s)\nu(s)}$ is the doubly symmetric traceless part of $\widetilde X_{\mu_1\cdots\mu_s\nu_1\cdots\nu_s}$ 
\begin{equation}
 \widetilde X_{\mu_1\cdots\mu_s\nu_1\cdots\nu_s}=\gamma_{\mu_1\nu_1}\cdots \gamma_{\mu_s\nu_s}.
\end{equation}
Here $\gamma_{\mu\nu}$ has been defined as
\begin{align}
 \gamma_{\mu\nu}=\eta_{\mu\nu}-\frac1{2}[{n_\mu(\bm k)\bar n_\nu(\bm k)+\bar n_\mu(\bm k) n_\nu(\bm k)}]=\gamma_{AB}Y_\mu^AY_\nu^B(\Omega_k),
\end{align}
with
\begin{align}
 n_\mu(\bm k)=(-1,n_i(\bm k)),\quad \bar{n}_\mu(\bm k)=(1,n_i(\bm k)),\quad n_i(\bm k)=\frac{k_i}{|\bm k|}.
\end{align}
In the context, our polarization tensor satisfies
\begin{align}
 \varepsilon'_{\mu(s-2)}=0,\qquad k^{\nu}\varepsilon_{\nu\mu(s-1)}=0,\qquad \varepsilon_{r\mu(s-1)}=0,
\end{align}
which imply the corresponding properties of \eqref{comp}. The property of being symmetric and traceless is obvious due to the construction of $X_{\mu(s)\nu(s)}$, while the others are satisfied by the definition of $\gamma_{\mu\nu}$, namely $k^\mu \gamma_{\mu\nu}=0$ and
\begin{align}
 Y_r^A=Y^A_\mu\p_rx^\mu=Y_\mu^An^\mu=0 \quad\Rightarrow\quad \gamma_{r\nu}=\gamma_{AB}Y_r^AY_\nu^B=0.
\end{align}

As in the context, we impose the fall-off
\begin{align}
 f_{\mu(s)}(t,\bm x)&=\frac{F_{\mu(s)}(u,\Omega)}{r}+\mathcal{O}({r^{-2}}), 
\end{align}
which leads to
\begin{align}
 F_{\mu(s)}(u,\Omega)&=\sum_{\alpha}\sum_{\ell, m}\int_0^\infty d\omega d\Omega_k [\frac{\sqrt{\omega}}{4\sqrt{2}\pi^2i}\varepsilon_{\mu(s)}^{*\alpha}(\bm k)b_{\alpha,\bm k} e^{-i\omega u}Y_{\ell, m}(\Omega)Y^*_{\ell, m}(\Omega_k)+\text{h.c.}]\nn\\
 &=\int_0^\infty \frac{d\omega}{\sqrt{4\pi\omega}} \sum_{\ell,m}[c_{\mu(s);\omega,\ell,m}e^{-i\omega u}Y_{\ell,m}(\Omega)+\text{h.c.}],
\end{align}
with
\begin{subequations}\label{cb}
\begin{align}
 c_{\mu(s);\omega,\ell,m}&=\frac{\omega}{(2\pi)^{3/2}i}\int d\Omega_k \sum_{\alpha}\varepsilon_{\mu(s)}^{*\alpha}(\bm k)b_{\alpha,\bm k}Y_{\ell,m}^*(\Omega_k),\\
 c_{\mu(s);\omega,\ell,m}^\dagger&=\frac{i\omega}{(2\pi)^{3/2}}\int d\Omega_k \sum_{\alpha}\varepsilon_{\mu(s)}^{\alpha}(\bm k)b^\dagger_{\alpha,\bm k}Y_{\ell,m}(\Omega_k).
\end{align}
\end{subequations}
Converting to retarded frame, we obtain
\begin{align}
 F_{A(s)}(u,\Omega)&=F_{\mu(s)}(u,\Omega)(-Y^{\mu_1}_{A_1})\cdots (-Y^{\mu_s}_{A_s})\nn\\
 &=\int_0^\infty \frac{d\omega}{\sqrt{4\pi\omega}}\sum_{\ell m}[(-1)^sc_{i(s);\omega,\ell,m}Y^{i(s)}_{A(s)}Y_{\ell,m}(\Omega)e^{-i\omega u}+\text{h.c.}].
\end{align}

It is straightforward to compute the commutation relation between boundary creation and annihilation operators
\begin{align}
 [c_{i(s);\omega,\ell,m},c^\dagger_{i'(s);\omega',\ell',m}]&=\frac{\omega\omega'}{(2\pi)^{3}}\int d\Omega_kd\Omega_k' \varepsilon_{i(s)}^{*\alpha}(\bm k)Y_{\ell,m}^*(\Omega_k) \varepsilon_{i'(s)}^{\alpha'}(\bm k')Y_{\ell',m'}^*(\Omega_k')[b_{\alpha,\bm k},b_{\alpha',\bm k'}^\dagger]\nn\\
 &=\delta(\omega-\omega')\int d\Omega_k X_{i(s)i'(s)} Y_{\ell,m}^*(\Omega_k)Y_{\ell',m'}(\Omega_k).
\end{align}
Now we are prepared to calculate the fundamental commutator
\begin{align}
 &[F_{A(s)}(u,\Omega),F_{B(s)}(u',\Omega')]\nn\\
 &=\int_0^\infty \frac{d\omega}{\sqrt{4\pi\omega}}\frac{d\omega'}{\sqrt{4\pi\omega'}}[Y^{i(s)}_{A(s)}Y^{i'(s)}_{B(s)}\nn\\
 &\quad \times \sum_{\ell m}Y_{\ell,m}(\Omega)e^{-i\omega u}\sum_{\ell' m'}Y^*_{\ell',m'}(\Omega')e^{i\omega' u'}[c_{i(s);\omega,\ell,m},c^\dagger_{i'(s);\omega',\ell',m'}]+\text{h.c.}]\nn\\
 &=\frac{i}{2}\alpha(u-u')\delta(\Omega-\Omega')Y^{i(s)}_{A(s)}Y^{i'(s)}_{B(s)}X_{i(s)i'(s)}\nn\\
 &=\frac{i}{2}X_{A(s)B(s)}\alpha(u-u')\delta(\Omega-\Omega'),
\end{align}
which agrees with our previous result from boundary symplectic form.

We can also use the mode expansion to derive the antipodal matching conditions
\begin{align}
 F^+_{\mu(s)}(\omega,\Omega)=-F_{\mu(s)}^-(\omega,\Omega^P),\qquad F_{\mu(s)}^{+(2)}(\omega,\Omega)=F_{\mu(s)}^{-(2)}(\omega,\Omega^P)
\end{align}
up to the first two orders, where $\Omega^P=(\pi-\theta,\pi+\phi)$ is the antipodal point of $\Omega=(\theta,\phi)$ on the sphere, and $+$ ($-$) denotes fields at $\mathcal{I}^+$ ($\mathcal{I}^-$). This result has also been checked using Green's function for retarded and advanced solutions of the wave equation with source.

\subsection{Polarization tensors}
In this subsection, we discuss the polarization tensors in HS theory. 
\paragraph{Spin one and special momentum.} 
For simplicity, we consider the case of $s=1$, and take a special momentum $k_\mu=|\bm k|(1,0,0,1)$ which is followed by
\begin{align}
 \gamma_\mn=\begin{pmatrix}
 0 &&&\\ &1&&\\ &&1&\\ &&& 0
 \end{pmatrix},\qquad
 \bar\ep_\mn=\begin{pmatrix}
 0 &&&\\ &0&1&\\ &-1&0&\\ &&& 0
 \end{pmatrix}.
\end{align} 
We need the polarization vectors to satisfy the orthogonality and completeness relations
\begin{align}
 &\gamma^{\mu\mu'}\varepsilon_{\mu}^{*\alpha}(\bm k)\varepsilon_{\mu'}^{\alpha'}(\bm k)=\delta^{\a\a'},
 \\ 
 &\sum_{\alpha,\beta}\varepsilon_{\mu}^{*\alpha}(\bm k)\delta_{\alpha,\beta}\varepsilon_{\nu}^{\beta}(\bm k)=\gamma_{\mu\nu},
\end{align}
and the transverse condition
\begin{align}
 k^\mu\varepsilon^\a_\mu(\bm k)=0.
\end{align}
A natural choice is 
\begin{align}
 \varepsilon^{\rm R}_\mu=\frac{1}{\sqrt2}(0,1,i,0),\qquad \varepsilon^{\rm L}_\mu=\frac{1}{\sqrt2}(0,1,-i,0),
\end{align}
which also satisfy the condition
\begin{align}
 \bar\epsilon^{\nu\mu}(\Omega_k)\varepsilon_{\mu}^{*\a}(\bm k)\varepsilon_{\nu}^{\b}(\bm k)=i\sigma_3^{\ab},\label{eqe21}
\end{align}
and thus agree with \eqref{eq628}.

\paragraph{General momentum.}
For a general momentum $k_\mu$, the construction of the polarization vectors may be rather complicated. However, we find that the properties that they need to satisfy happen to be the ones of $Y^A_\mu$, namely
\begin{align}
 &Y_\mu^AY^B_\nu\gamma_{AB}= \gamma_\mn,\\
 &\gamma^\mn Y^A_\mu Y^B_\n=\eta^\mn Y^A_\mu Y^B_\n=\g^{AB},
\end{align}
and $n^\mu Y^A_\mu=0$, except for \eqref{eqe21}. Therefore, one can introduce the vielbeins 
\begin{align}
 e_1^A=\frac{1}{\sqrt2}\begin{pmatrix}
 1\\ \sin^{-1}\th
 \end{pmatrix},\qquad e_2^A=\frac{1}{\sqrt2}
 \begin{pmatrix}
 1\\ -\sin^{-1}\th
 \end{pmatrix}
\end{align}
such that
\begin{align}
 \delta^{\alpha\beta}e_\a^Ae_\b^B=\g^{AB}.
\end{align}
It follows that the polarization vectors can be expressed as
\begin{align}
 \varepsilon_{\mu}^{A}=\varepsilon_{\mu}^{\alpha}e_\a^A\equiv Y^A_\mu.\label{cho}
\end{align}
One can invert the relation to obtain
\begin{align}
 \varepsilon_\mu^\a=Y^A_\mu e_A^\a,\qquad e_A^\a=e_\b^B\g_{AB}\delta^{\ab}.
\end{align}

With the choice \eqref{cho}, we find
\begin{align}
 &\bar\epsilon^{\mu\nu}Y_{\mu}^AY_{\nu}^B=\epsilon^{CD}Y^\mu_CY^\nu_DY_{\mu}^AY_{\nu}^B=\epsilon^{AB}=-\epsilon^{\ab}e_\a^Ae_\b^B\\
 \Leftrightarrow\quad &\bar\epsilon^{\nu\mu}\varepsilon_\mu^{*\a}\varepsilon_\nu^\b=\epsilon^{\ab},
\end{align} which is not the last property \eqref{eqe21} superficially. However, 
one can combine the polarization vectors to get
\begin{align}
 \varepsilon_\mu^{\rm R}=\frac1{\sqrt 2}(\varepsilon_\mu^1+i\varepsilon_\mu^2),\qquad \varepsilon_\mu^{\rm L}=\frac1{\sqrt 2}(\varepsilon_\mu^1-i\varepsilon_\mu^2),\label{e30}
\end{align}
which satisfy
\begin{align}
 \bar\epsilon^{\nu\mu}\varepsilon_\mu^{*\a}\varepsilon_\nu^{\b}=i\s_3^\ab,\qquad \a,\b={\rm R,\, L},
\end{align}
as we want.

\paragraph{General spin.}
The key point to derive the HS polarization tensors is noting that \eqref{e30} can be rewritten as
\begin{subequations}
\begin{align}
 &\varepsilon_\mu^{\rm R}=\frac{1}{2}[(1+i)Y^\th_\mu+(1-i)\sin\th Y^\phi_\mu]=\frac{1}{2}(1+i) (Y_\mu+i\widetilde{Y}_\mu)^\th\equiv\frac{1}{2}(1+i)\cy_\mu,\\
 &\varepsilon_\mu^{\rm L}=\frac{1}{2}[(1-i)Y^\th_\mu+(1+i)\sin\th Y^\phi_\mu]=\frac{1}{2}(1-i) (Y_\mu-i\widetilde{Y}_\mu)^\th\equiv\frac{1}{2}(1-i)\bar{\cy}_\mu,
\end{align}
\end{subequations}
where we have defined 
\begin{align}
 \widetilde{Y}_{\mu}^A=Y_{\mu}^B\epsilon_{B}{}^A=(0,\widetilde{Y}_i^A),
\end{align}
which can also be interpreted as a Hodge dual
\begin{align}
 \widetilde{Y}_{\mu}^A= \widetilde{Y}_{0\mu}^A=(0,\frac{1}{2}\epsilon_{ijk}Y_{jk}^A),\qquad \widetilde{Y}^A_{\mn}=\frac{1}{2}\epsilon_{\mn\rs}Y^{\rs A}.
\end{align}
One can easily find 
\begin{align}
 \cy_\mu=(Y_\mu+i\widetilde{Y}_\mu)^\th=
 \begin{pmatrix}
 0\\
 -\cos\theta\cos\phi-i\sin\phi\\
 -\cos\theta\sin\phi+i\cos\phi\\
 \sin\th
 \end{pmatrix},\label{3.123}
\end{align}
and its complex conjugate $\bar\cy_\mu$, which agree with the expression in the literature, such as \cite{Elvang:2015rqa}. Now we can construct the polarization tensors for the HS theory
\begin{subequations}
\begin{align}
 &\varepsilon_{\mu(s)}^{\rm R}=\varepsilon_{\mu_1}^{\rm R}\cdots \varepsilon_{\mu_s}^{\rm R}=\frac{(1+i)^s}{2^s}\cy_{\mu_1}\cdots \cy_{\mu_s}=\frac{(1+i)^s}{2^s}\cy_{\mu(s)},\\
 &\varepsilon_{\mu(s)}^{\rm L}=\varepsilon_{\mu_1}^{\rm L}\cdots \varepsilon_{\mu_s}^{\rm L}=\frac{(1-i)^s}{2^s}\bar{\cy}_{\mu_1}\cdots \bar{\cy}_{\mu_s}=\frac{(1-i)^s}{2^s}\bar{\cy}_{\mu(s)}.
\end{align}
\end{subequations}
A nice property is that these expressions are automatically symmetric and traceless, since we have 
\begin{subequations}
\begin{align}
 &\eta^\mn(Y_\mu+i\widetilde{Y}_\mu)^A(Y_\nu+i\widetilde{Y}_\nu)^B=\g^\mn(Y_\mu+i\widetilde{Y}_\mu)^A(Y_\nu+i\widetilde{Y}_\nu)^B=0,\\
 &\eta^\mn(Y_\mu-i\widetilde{Y}_\mu)^A(Y_\nu-i\widetilde{Y}_\nu)^B=\g^\mn(Y_\mu-i\widetilde{Y}_\mu)^A(Y_\nu-i\widetilde{Y}_\nu)^B=0,
\end{align}
\end{subequations}
due to the identities \(n^\mu Y_\mu^A=n^\mu \widetilde{Y}_\mu^A=
\bar n^\mu Y_\mu^A=\bar n^\mu \widetilde{Y}_\mu^A=0\) and
\begin{align}
 Y^A\cdot Y^B=\widetilde{Y}^A\cdot\widetilde{Y}^B=\g^{AB},\qquad Y^A\cdot \widetilde{Y}^B=\epsilon^{AB}.
\end{align}

Now we need to check the orthogonality and completeness relations, as well as \eqref{eq628}. The orthogonality relation is straightforward
\begin{align}
 X^{\m(s)\n(s)}\varepsilon_{\mu(s)}^{*\a}\varepsilon_{\nu(s)}^{\a'}=\delta^{\a\a'},
\end{align}
since we have
\begin{align}
 \g^\mn\bar\cy_{\mu}\cy_\nu=2,\quad \gamma^{\mu\nu}\cy_{\mu}\cy_\nu=\gamma^{\mu\nu}\bar\cy_{\mu}\bar\cy_\nu=0.\label{yyb}
\end{align}

The completeness relation reads
\begin{align}
 \varepsilon_{\mu(s)}^{*\rm R}\varepsilon_{\nu(s)}^{\rm R}+\varepsilon_{\mu(s)}^{*\rm L}\varepsilon_{\nu(s)}^{\rm L}=2^{-s}[\bar\cy_{\mu_1}\cdots \bar\cy_{\mu_s}\cy_{\nu_1}\cdots \cy_{\nu_s}+\c.c.]=X_{\mu(s)\nu(s)},\label{3.129}
\end{align}
which is a bit difficult to prove. For $s=1$, we know that it is satisfied
\begin{align}
 \bar\cy_\mu\cy_\nu+\c.c.=2\g_{\mn}.
\end{align}
For $s=2$, we find
\begin{align}
 X_{\mu(2)\nu(2)}&=\frac12(\g_{\mu_1\nu_1}\g_{\mu_2\nu_2}+\g_{\mu_1\nu_2}\g_{\mu_2\nu_1}-\g_{\mu_1\m_2}\g_{\nu_1\nu_2})\nn\\
 &=\frac{1}{2}\times\frac{1}{4}\Big[(\bar\cy_{\mu_1}\cy_{\nu_1}+\c.c.)(\bar\cy_{\mu_2}\cy_{\nu_2}+\c.c.)+(\bar\cy_{\mu_1}\cy_{\nu_2}+\c.c.)(\bar\cy_{\mu_2}\cy_{\nu_1}+\c.c.)\nn\\
 &\quad -(\bar\cy_{\mu_1}\cy_{\mu_2}+\c.c.)(\bar\cy_{\nu_1}\cy_{\nu_2}+\c.c.)\Big]\nn\\
 &=\frac{1}{4}[\bar\cy_{\mu_1}\bar\cy_{\mu_2}\cy_{\nu_1}\cy_{\nu_2}+\c.c.].
\end{align}
In general, we find that the right-hand side of the completeness relation is
\begin{align}
 {\rm DST}[\g_{\mu_1\nu_1}\cdots\g_{\mu_s\nu_s}]&=2^{-s}{\rm DST}[(\bar\cy_{\mu_1}\cy_{\nu_1}+\c.c.)\cdots (\bar\cy_{\mu_s}\cy_{\nu_s}+\c.c.)],
\end{align}
which contains a same number of $\bar\cy$ and $\cy$. The notation ``DST[$\cdots$]'' represents the doubly symmetric traceless part of the expression inside the square brackets. To be doubly symmetric traceless, we must have
\begin{align}
 &{\rm DST}[(\bar\cy_{\mu_1}\cy_{\nu_1}+\c.c.)\cdots (\bar\cy_{\mu_s}\cy_{\nu_s}+\c.c.)]\propto \bar\cy_{\mu_1}\cdots \bar\cy_{\mu_s}\cy_{\nu_1}\cdots \cy_{\nu_s}+\c.c.,
\end{align}
due to the relations \eqref{yyb}. 
Then the overall coefficient is easily to be determined.

At last, we can check
\begin{align}
 Q^{\mu(s)\mu'(s)}\varepsilon_{\mu(s)}^{*\alpha}\varepsilon_{\mu'(s)}^{\alpha'}=i\s_3^{\a\a'},\qquad \a,\a'={\rm R,L},
\end{align}
where we need to use 
\begin{align}
 Y^\mu_A\bar\cy_\m=Y^\mu_A(Y_\mu-i\widetilde{Y}_\mu)^\th=\g_A^\th-i\ep_A{}^\th,
\end{align}
and thus
\begin{align}
 &\frac{1}{2}\bar{\ep}^{\nu\mu}\bar\cy_{\mu}\cy_{\nu}=\frac{1}{2}\ep^{AB}Y^\nu_A Y^\mu_B\bar\cy_{\mu}\cy_{\nu}=i.
\end{align}

\bibliography{refs}

\providecommand{\href}[2]{#2}\begingroup\raggedright\begin{thebibliography}{10}

\bibitem{Une}
J.~M. L\'evy-Leblond, ``{Une nouvelle limite non-relativiste du groupe de
  Poincar\'e},'' {\em Ann. Inst. H Poincar\'e} {\bf 3} (1965), no.~1, 1--12.

\bibitem{Gupta1966OnAA}
N.~Gupta, ``On an analogue of the galilei group,'' {\em Nuovo Cimento Della
  Societa Italiana Di Fisica A-nuclei Particles and Fields} {\bf 44} (1966)
  512--517.

\bibitem{Duval_2014a}
C.~Duval, G.~W. Gibbons, and P.~A. Horvathy, ``Conformal carroll groups and
  {BMS} symmetry,'' {\em Classical and Quantum Gravity} {\bf 31} (apr, 2014)
  092001.

\bibitem{Duval_2014b}
C.~Duval, G.~W. Gibbons, and P.~A. Horvathy, ``Conformal carroll groups,'' {\em
  Journal of Physics A: Mathematical and Theoretical} {\bf 47} (aug, 2014)
  335204.

\bibitem{Duval:2014uoa}
C.~Duval, G.~W. Gibbons, P.~A. Horvathy, and P.~M. Zhang, ``{Carroll versus
  Newton and Galilei: two dual non-Einsteinian concepts of time},'' {\em Class.
  Quant. Grav.} {\bf 31} (2014) 085016,
  \href{http://www.arXiv.org/abs/1402.0657}{{\tt 1402.0657}}.

\bibitem{Bondi:1962px}
H.~Bondi, M.~G.~J. van~der Burg, and A.~W.~K. Metzner, ``{Gravitational waves
  in general relativity. 7. Waves from axisymmetric isolated systems},'' {\em
  Proc. Roy. Soc. Lond. A} {\bf 269} (1962) 21--52.

\bibitem{Sachs:1962wk}
R.~K. Sachs, ``{Gravitational waves in general relativity. 8. Waves in
  asymptotically flat space-times},'' {\em Proc. Roy. Soc. Lond. A} {\bf 270}
  (1962) 103--126.

\bibitem{Barnich:2009se}
G.~Barnich and C.~Troessaert, ``{Symmetries of asymptotically flat 4
  dimensional spacetimes at null infinity revisited},'' {\em Phys. Rev. Lett.}
  {\bf 105} (2010) 111103, \href{http://www.arXiv.org/abs/0909.2617}{{\tt
  0909.2617}}.

\bibitem{Barnich:2010eb}
G.~Barnich and C.~Troessaert, ``{Aspects of the BMS/CFT correspondence},'' {\em
  JHEP} {\bf 05} (2010) 062, \href{http://www.arXiv.org/abs/1001.1541}{{\tt
  1001.1541}}.

\bibitem{Campiglia:2014yka}
M.~Campiglia and A.~Laddha, ``{Asymptotic symmetries and subleading soft
  graviton theorem},'' {\em Phys. Rev. D} {\bf 90} (2014), no.~12, 124028,
  \href{http://www.arXiv.org/abs/1408.2228}{{\tt 1408.2228}}.

\bibitem{Campiglia:2015yka}
M.~Campiglia and A.~Laddha, ``{New symmetries for the Gravitational
  S-matrix},'' {\em JHEP} {\bf 04} (2015) 076,
  \href{http://www.arXiv.org/abs/1502.02318}{{\tt 1502.02318}}.

\bibitem{Freidel:2021fxf}
L.~Freidel, R.~Oliveri, D.~Pranzetti, and S.~Speziale, ``{The Weyl BMS group
  and Einstein\textquoteright{}s equations},'' {\em JHEP} {\bf 07} (2021) 170,
  \href{http://www.arXiv.org/abs/2104.05793}{{\tt 2104.05793}}.

\bibitem{Liu:2022mne}
W.-B. Liu and J.~Long, ``{Symmetry group at future null infinity: Scalar
  theory},'' {\em Phys. Rev. D} {\bf 107} (2023), no.~12, 126002,
  \href{http://www.arXiv.org/abs/2210.00516}{{\tt 2210.00516}}.

\bibitem{Liu:2023qtr}
W.-B. Liu and J.~Long, ``{Symmetry group at future null infinity II: Vector
  theory},'' {\em JHEP} {\bf 07} (2023) 152,
  \href{http://www.arXiv.org/abs/2304.08347}{{\tt 2304.08347}}.

\bibitem{Liu:2023gwa}
W.-B. Liu and J.~Long, ``{Symmetry group at future null infinity III:
  Gravitational theory},'' {\em JHEP} {\bf 10} (2023) 117,
  \href{http://www.arXiv.org/abs/2307.01068}{{\tt 2307.01068}}.

\bibitem{Li:2023xrr}
A.~Li, W.-B. Liu, J.~Long, and R.-Z. Yu, ``{Quantum flux operators for
  Carrollian diffeomorphism in general dimensions},'' {\em JHEP} {\bf 11}
  (2023) 140, \href{http://www.arXiv.org/abs/2309.16572}{{\tt 2309.16572}}.

\bibitem{1964PhRv..135.1049W}
S.~{Weinberg}, ``{Photons and Gravitons in S-Matrix Theory: Derivation of
  Charge Conservation and Equality of Gravitational and Inertial Mass},'' {\em
  Physical Review} {\bf 135} (Aug., 1964) 1049--1056.

\bibitem{Grisaru:1976vm}
M.~T. Grisaru, H.~N. Pendleton, and P.~van Nieuwenhuizen, ``{Supergravity and
  the S Matrix},'' {\em Phys. Rev. D} {\bf 15} (1977) 996.

\bibitem{Aragone:1979hx}
C.~Aragone and S.~Deser, ``{Consistency Problems of Hypergravity},'' {\em Phys.
  Lett. B} {\bf 86} (1979) 161--163.

\bibitem{Weinberg:1980kq}
S.~Weinberg and E.~Witten, ``{Limits on Massless Particles},'' {\em Phys. Lett.
  B} {\bf 96} (1980) 59--62.

\bibitem{Porrati:2008rm}
M.~Porrati, ``{Universal Limits on Massless High-Spin Particles},'' {\em Phys.
  Rev. D} {\bf 78} (2008) 065016,
  \href{http://www.arXiv.org/abs/0804.4672}{{\tt 0804.4672}}.

\bibitem{Campoleoni:2017mbt}
A.~Campoleoni, D.~Francia, and C.~Heissenberg, ``{On higher-spin
  supertranslations and superrotations},'' {\em JHEP} {\bf 05} (2017) 120,
  \href{http://www.arXiv.org/abs/1703.01351}{{\tt 1703.01351}}.

\bibitem{Campoleoni:2017qot}
A.~Campoleoni, D.~Francia, and C.~Heissenberg, ``{Asymptotic Charges at Null
  Infinity in Any Dimension},'' {\em Universe} {\bf 4} (2018), no.~3, 47,
  \href{http://www.arXiv.org/abs/1712.09591}{{\tt 1712.09591}}.

\bibitem{Campoleoni:2020ejn}
A.~Campoleoni, D.~Francia, and C.~Heissenberg, ``{On asymptotic symmetries in
  higher dimensions for any spin},'' {\em JHEP} {\bf 12} (2020) 129,
  \href{http://www.arXiv.org/abs/2011.04420}{{\tt 2011.04420}}.

\bibitem{Bekaert:2022oeh}
X.~Bekaert, A.~Campoleoni, and S.~Pekar, ``{Carrollian conformal scalar as
  flat-space singleton},'' {\em Phys. Lett. B} {\bf 838} (2023) 137734,
  \href{http://www.arXiv.org/abs/2211.16498}{{\tt 2211.16498}}.

\bibitem{Dirac:1931kp}
P.~A.~M. Dirac, ``{Quantised singularities in the electromagnetic field,},''
  {\em Proc. Roy. Soc. Lond. A} {\bf 133} (1931), no.~821, 60--72.

\bibitem{tHooft:1974kcl}
G.~'t~Hooft, ``{Magnetic Monopoles in Unified Gauge Theories},'' {\em Nucl.
  Phys. B} {\bf 79} (1974) 276--284.

\bibitem{Polyakov:1974ek}
A.~M. Polyakov, ``{Particle Spectrum in Quantum Field Theory},'' {\em JETP
  Lett.} {\bf 20} (1974) 194--195.

\bibitem{1976PhRvD..13.1592D}
S.~{Deser} and C.~{Teitelboim}, ``{Duality transformations of Abelian and
  non-Abelian gauge fields},'' {\em Phys Review D} {\bf 13} (Mar., 1976)
  1592--1597.

\bibitem{Montonen:1977sn}
C.~Montonen and D.~I. Olive, ``{Magnetic Monopoles as Gauge Particles?},'' {\em
  Phys. Lett. B} {\bf 72} (1977) 117--120.

\bibitem{Nepomechie:1984wu}
R.~I. Nepomechie, ``{Magnetic Monopoles from Antisymmetric Tensor Gauge
  Fields},'' {\em Phys. Rev. D} {\bf 31} (1985) 1921.

\bibitem{Teitelboim:1985ya}
C.~Teitelboim, ``{Gauge Invariance for Extended Objects},'' {\em Phys. Lett. B}
  {\bf 167} (1986) 63--68.

\bibitem{Teitelboim:1985yc}
C.~Teitelboim, ``{Monopoles of Higher Rank},'' {\em Phys. Lett. B} {\bf 167}
  (1986) 69--72.

\bibitem{Garcia-Compean:1998ipq}
H.~Garcia-Compean, O.~Obregon, and C.~Ramirez, ``{Gravitational duality in
  MacDowell-Mansouri gauge theory},'' {\em Phys. Rev. D} {\bf 58} (1998)
  104012, \href{http://www.arXiv.org/abs/hep-th/9802063}{{\tt hep-th/9802063}}.

\bibitem{Nieto:1999pn}
J.~A. Nieto, ``{S duality for linearized gravity},'' {\em Phys. Lett. A} {\bf
  262} (1999) 274--281, \href{http://www.arXiv.org/abs/hep-th/9910049}{{\tt
  hep-th/9910049}}.

\bibitem{Hull:2000zn}
C.~M. Hull, ``{Strongly coupled gravity and duality},'' {\em Nucl. Phys. B}
  {\bf 583} (2000) 237--259,
  \href{http://www.arXiv.org/abs/hep-th/0004195}{{\tt hep-th/0004195}}.

\bibitem{Casini:2003kf}
H.~Casini, R.~Montemayor, and L.~F. Urrutia, ``{Duality for symmetric second
  rank tensors. 2. The Linearized gravitational field},'' {\em Phys. Rev. D}
  {\bf 68} (2003) 065011, \href{http://www.arXiv.org/abs/hep-th/0304228}{{\tt
  hep-th/0304228}}.

\bibitem{Henneaux:2004jw}
M.~Henneaux and C.~Teitelboim, ``{Duality in linearized gravity},'' {\em Phys.
  Rev. D} {\bf 71} (2005) 024018,
  \href{http://www.arXiv.org/abs/gr-qc/0408101}{{\tt gr-qc/0408101}}.

\bibitem{Godazgar:2018qpq}
H.~Godazgar, M.~Godazgar, and C.~N. Pope, ``{New dual gravitational charges},''
  {\em Phys. Rev. D} {\bf 99} (2019), no.~2, 024013,
  \href{http://www.arXiv.org/abs/1812.01641}{{\tt 1812.01641}}.

\bibitem{Oliveri:2020xls}
R.~Oliveri and S.~Speziale, ``{A note on dual gravitational charges},'' {\em
  JHEP} {\bf 12} (2020) 079, \href{http://www.arXiv.org/abs/2010.01111}{{\tt
  2010.01111}}.

\bibitem{Seraj:2022qyt}
A.~Seraj and B.~Oblak, ``{Precession Caused by Gravitational Waves},'' {\em
  Phys. Rev. Lett.} {\bf 129} (2022), no.~6, 061101,
  \href{http://www.arXiv.org/abs/2203.16216}{{\tt 2203.16216}}.

\bibitem{Witten:1978mh}
E.~Witten and D.~I. Olive, ``{Supersymmetry Algebras That Include Topological
  Charges},'' {\em Phys. Lett. B} {\bf 78} (1978) 97--101.

\bibitem{Osborn:1979tq}
H.~Osborn, ``{Topological Charges for N=4 Supersymmetric Gauge Theories and
  Monopoles of Spin 1},'' {\em Phys. Lett. B} {\bf 83} (1979) 321--326.

\bibitem{Seiberg:1994rs}
N.~Seiberg and E.~Witten, ``{Electric - magnetic duality, monopole
  condensation, and confinement in N=2 supersymmetric Yang-Mills theory},''
  {\em Nucl. Phys. B} {\bf 426} (1994) 19--52,
  \href{http://www.arXiv.org/abs/hep-th/9407087}{{\tt hep-th/9407087}}.
  [Erratum: Nucl.Phys.B 430, 485--486 (1994)].

\bibitem{Intriligator:1995au}
K.~A. Intriligator and N.~Seiberg, ``{Lectures on supersymmetric gauge theories
  and electric-magnetic duality},'' {\em Nucl. Phys. B Proc. Suppl.} {\bf 45BC}
  (1996) 1--28, \href{http://www.arXiv.org/abs/hep-th/9509066}{{\tt
  hep-th/9509066}}.

\bibitem{Francia:2004lbf}
D.~Francia and C.~M. Hull, ``{Higher-spin gauge fields and duality},'' in {\em
  {1st Solvay Workshop on Higher Spin Gauge Theories}}, pp.~35--48.
\newblock 2004.
\newblock \href{http://www.arXiv.org/abs/hep-th/0501236}{{\tt hep-th/0501236}}.

\bibitem{Bekaert:2002dt}
X.~Bekaert and N.~Boulanger, ``{Tensor gauge fields in arbitrary
  representations of GL(D,R): Duality and Poincare lemma},'' {\em Commun. Math.
  Phys.} {\bf 245} (2004) 27--67,
  \href{http://www.arXiv.org/abs/hep-th/0208058}{{\tt hep-th/0208058}}.

\bibitem{deMedeiros:2003osq}
P.~de~Medeiros and C.~Hull, ``{Geometric second order field equations for
  general tensor gauge fields},'' {\em JHEP} {\bf 05} (2003) 019,
  \href{http://www.arXiv.org/abs/hep-th/0303036}{{\tt hep-th/0303036}}.

\bibitem{Bekaert:2003az}
X.~Bekaert and N.~Boulanger, ``{On geometric equations and duality for free
  higher spins},'' {\em Phys. Lett. B} {\bf 561} (2003) 183--190,
  \href{http://www.arXiv.org/abs/hep-th/0301243}{{\tt hep-th/0301243}}.

\bibitem{Boulanger:2003vs}
N.~Boulanger, S.~Cnockaert, and M.~Henneaux, ``{A note on spin s duality},''
  {\em JHEP} {\bf 06} (2003) 060,
  \href{http://www.arXiv.org/abs/hep-th/0306023}{{\tt hep-th/0306023}}.

\bibitem{Deser:2004xt}
S.~Deser and D.~Seminara, ``{Duality invariance of all free bosonic and
  fermionic gauge fields},'' {\em Phys. Lett. B} {\bf 607} (2005) 317--319,
  \href{http://www.arXiv.org/abs/hep-th/0411169}{{\tt hep-th/0411169}}.

\bibitem{Bunster:2006rt}
C.~W. Bunster, S.~Cnockaert, M.~Henneaux, and R.~Portugues, ``{Monopoles for
  gravitation and for higher spin fields},'' {\em Phys. Rev. D} {\bf 73} (2006)
  105014, \href{http://www.arXiv.org/abs/hep-th/0601222}{{\tt hep-th/0601222}}.

\bibitem{Henneaux:2016zlu}
M.~Henneaux, S.~H\"ortner, and A.~Leonard, ``{Twisted self-duality for higher
  spin gauge fields and prepotentials},'' {\em Phys. Rev. D} {\bf 94} (2016),
  no.~10, 105027, \href{http://www.arXiv.org/abs/1609.04461}{{\tt 1609.04461}}.
  [Erratum: Phys.Rev.D 97, 049901 (2018)].

\bibitem{Vasiliev:1990en}
M.~A. Vasiliev, ``{Consistent equation for interacting gauge fields of all
  spins in (3+1)-dimensions},'' {\em Phys. Lett. B} {\bf 243} (1990) 378--382.

\bibitem{Vasiliev:1992av}
M.~A. Vasiliev, ``{More on equations of motion for interacting massless fields
  of all spins in (3+1)-dimensions},'' {\em Phys. Lett. B} {\bf 285} (1992)
  225--234.

\bibitem{Vasiliev:2003ev}
M.~A. Vasiliev, ``{Nonlinear equations for symmetric massless higher spin
  fields in (A)dS(d)},'' {\em Phys. Lett. B} {\bf 567} (2003) 139--151,
  \href{http://www.arXiv.org/abs/hep-th/0304049}{{\tt hep-th/0304049}}.

\bibitem{Vasiliev:2003cph}
M.~A. Vasiliev, ``{Higher spin gauge theories in various dimensions},'' {\em
  PoS} {\bf JHW2003} (2003) 003,
  \href{http://www.arXiv.org/abs/hep-th/0401177}{{\tt hep-th/0401177}}.

\bibitem{Metsaev:1991mt}
R.~R. Metsaev, ``{Poincare invariant dynamics of massless higher spins: Fourth
  order analysis on mass shell},'' {\em Mod. Phys. Lett. A} {\bf 6} (1991)
  359--367.

\bibitem{Metsaev:1991nb}
R.~R. Metsaev, ``{S matrix approach to massless higher spins theory. 2: The
  Case of internal symmetry},'' {\em Mod. Phys. Lett. A} {\bf 6} (1991)
  2411--2421.

\bibitem{Ponomarev:2016lrm}
D.~Ponomarev and E.~D. Skvortsov, ``{Light-Front Higher-Spin Theories in Flat
  Space},'' {\em J. Phys. A} {\bf 50} (2017), no.~9, 095401,
  \href{http://www.arXiv.org/abs/1609.04655}{{\tt 1609.04655}}.

\bibitem{Skvortsov:2018jea}
E.~D. Skvortsov, T.~Tran, and M.~Tsulaia, ``{Quantum Chiral Higher Spin
  Gravity},'' {\em Phys. Rev. Lett.} {\bf 121} (2018), no.~3, 031601,
  \href{http://www.arXiv.org/abs/1805.00048}{{\tt 1805.00048}}.

\bibitem{Skvortsov:2020wtf}
E.~Skvortsov, T.~Tran, and M.~Tsulaia, ``{More on Quantum Chiral Higher Spin
  Gravity},'' {\em Phys. Rev. D} {\bf 101} (2020), no.~10, 106001,
  \href{http://www.arXiv.org/abs/2002.08487}{{\tt 2002.08487}}.

\bibitem{Krasnov:2021nsq}
K.~Krasnov, E.~Skvortsov, and T.~Tran, ``{Actions for self-dual Higher Spin
  Gravities},'' {\em JHEP} {\bf 08} (2021) 076,
  \href{http://www.arXiv.org/abs/2105.12782}{{\tt 2105.12782}}.

\bibitem{Herfray:2022prf}
Y.~Herfray, K.~Krasnov, and E.~Skvortsov, ``{Higher-spin self-dual Yang-Mills
  and gravity from the twistor space},'' {\em JHEP} {\bf 01} (2023) 158,
  \href{http://www.arXiv.org/abs/2210.06209}{{\tt 2210.06209}}.

\bibitem{Adamo:2022lah}
T.~Adamo and T.~Tran, ``{Higher-spin Yang\textendash{}Mills, amplitudes and
  self-duality},'' {\em Lett. Math. Phys.} {\bf 113} (2023), no.~3, 50,
  \href{http://www.arXiv.org/abs/2210.07130}{{\tt 2210.07130}}.

\bibitem{Boulanger:2023prx}
N.~Boulanger, A.~Campoleoni, and S.~Pekar, ``{New higher-spin curvatures in
  flat space},'' {\em Phys. Rev. D} {\bf 108} (2023), no.~10, L101904,
  \href{http://www.arXiv.org/abs/2306.05367}{{\tt 2306.05367}}.

\bibitem{1978PhRvD..18.3624F}
C.~{Fronsdal}, ``{Massless fields with integer spin},'' {\em Phys.Rev.D} {\bf
  18} (Nov., 1978) 3624--3629.

\bibitem{Fang:1978wz}
J.~Fang and C.~Fronsdal, ``{Massless Fields with Half Integral Spin},'' {\em
  Phys. Rev. D} {\bf 18} (1978) 3630.

\bibitem{Fronsdal:1978vb}
C.~Fronsdal, ``{Singletons and Massless, Integral Spin Fields on de Sitter
  Space (Elementary Particles in a Curved Space. 7.},'' {\em Phys. Rev. D} {\bf
  20} (1979) 848--856.

\bibitem{Fang:1979hq}
J.~Fang and C.~Fronsdal, ``{Massless, Half Integer Spin Fields in De Sitter
  Space},'' {\em Phys. Rev. D} {\bf 22} (1980) 1361.

\bibitem{2018grav.book.....M}
C.~W. {Misner}, K.~S. {Thorne}, and J.~A. {Wheeler}, {\em {Gravitation}}.
\newblock {Princeton University Press}, 2018.

\bibitem{1980PhRvD..21..358D}
B.~{de Wit} and D.~Z. {Freedman}, ``{Systematics of higher-spin gauge
  fields},'' {\em Phys.Rev D} {\bf 21} (Jan., 1980) 358--367.

\bibitem{Berends:1985xx}
F.~A. Berends, G.~J.~H. Burgers, and H.~van Dam, ``{Explicit Construction of
  Conserved Currents for Massless Fields of Arbitrary Spin},'' {\em Nucl. Phys.
  B} {\bf 271} (1986) 429--441.

\bibitem{Anselmi:1999bb}
D.~Anselmi, ``{Higher spin current multiplets in operator product
  expansions},'' {\em Class. Quant. Grav.} {\bf 17} (2000) 1383--1400,
  \href{http://www.arXiv.org/abs/hep-th/9906167}{{\tt hep-th/9906167}}.

\bibitem{Konstein:2000bi}
S.~E. Konstein, M.~A. Vasiliev, and V.~N. Zaikin, ``{Conformal higher spin
  currents in any dimension and AdS / CFT correspondence},'' {\em JHEP} {\bf
  12} (2000) 018, \href{http://www.arXiv.org/abs/hep-th/0010239}{{\tt
  hep-th/0010239}}.

\bibitem{Gelfond:2006be}
O.~A. Gelfond, E.~D. Skvortsov, and M.~A. Vasiliev, ``{Higher spin conformal
  currents in Minkowski space},'' {\em Theor. Math. Phys.} {\bf 154} (2008)
  294--302, \href{http://www.arXiv.org/abs/hep-th/0601106}{{\tt
  hep-th/0601106}}.

\bibitem{Bekaert:2010hk}
X.~Bekaert and E.~Meunier, ``{Higher spin interactions with scalar matter on
  constant curvature spacetimes: conserved current and cubic coupling
  generating functions},'' {\em JHEP} {\bf 11} (2010) 116,
  \href{http://www.arXiv.org/abs/1007.4384}{{\tt 1007.4384}}.

\bibitem{Deser:2003rxd}
S.~Deser and A.~Waldron, ``{Stress and strain: T**mu nu of higher spin gauge
  fields},'' {\em PoS} {\bf jhw2003} (2003) 011,
  \href{http://www.arXiv.org/abs/hep-th/0403059}{{\tt hep-th/0403059}}.

\bibitem{1971ctf..book.....L}
L.~D. {Landau} and E.~M. {Lifshitz}, {\em {The classical theory of fields}}.
\newblock Springer Berlin, Heidelberg, 2012.

\bibitem{Smirnov:2013kba}
P.~A. Smirnov and M.~A. Vasiliev, ``{Gauge-noninvariant higher-spin currents in
  four-dimensional Minkowski space},'' {\em Theor. Math. Phys.} {\bf 181}
  (2014), no.~3, 1509--1521, \href{http://www.arXiv.org/abs/1312.6638}{{\tt
  1312.6638}}.

\bibitem{Campoleoni:2021blr}
A.~Campoleoni and S.~Pekar, ``{Carrollian and Galilean conformal higher-spin
  algebras in any dimensions},'' {\em JHEP} {\bf 02} (2022) 150,
  \href{http://www.arXiv.org/abs/2110.07794}{{\tt 2110.07794}}.

\bibitem{Bekaert:2022ipg}
X.~Bekaert and B.~Oblak, ``{Massless scalars and higher-spin BMS in any
  dimension},'' {\em JHEP} {\bf 11} (2022) 022,
  \href{http://www.arXiv.org/abs/2209.02253}{{\tt 2209.02253}}.

\bibitem{Bekaert:2006py}
X.~Bekaert and N.~Boulanger, ``{The unitary representations of the Poincar\'e
  group in any spacetime dimension},'' {\em SciPost Phys. Lect. Notes} {\bf 30}
  (2021) 1, \href{http://www.arXiv.org/abs/hep-th/0611263}{{\tt
  hep-th/0611263}}.

\bibitem{1965lgr..book.....T}
A.~{Trautmann}, F.~A.~E. {Pirani}, and H.~{Bondi}, {\em {Lectures on general
  relativity}}.
\newblock Prentice-Hall, Englewood Cliff, 1965.

\bibitem{Thorne:1980ru}
K.~S. Thorne, ``{Multipole Expansions of Gravitational Radiation},'' {\em Rev.
  Mod. Phys.} {\bf 52} (1980) 299--339.

\bibitem{Toth:2021cpx}
V.~T. Toth and S.~G. Turyshev, ``{Efficient trace-free decomposition of
  symmetric tensors of arbitrary rank},'' {\em Int. J. Geom. Meth. Mod. Phys.}
  {\bf 19} (2022), no.~13, 2250201,
  \href{http://www.arXiv.org/abs/2109.11743}{{\tt 2109.11743}}.

\bibitem{Elvang:2015rqa}
H.~Elvang and Y.-t. Huang, {\em {Scattering Amplitudes in Gauge Theory and
  Gravity}}.
\newblock Cambridge University Press, 4, 2015.

\end{thebibliography}\endgroup
\end{document}